\documentclass[aps,pra,showpacs,showkeys,onecolumn,notitlepage,groupedaddress]{revtex4-2}
\usepackage{graphicx}
\usepackage{multirow}
\usepackage{subcaption}
\usepackage{mwe}
\usepackage{braket}

\begin{document}

\title{Improving student understanding via interactive learning tutorial on quantum key distribution using entanglement} 

\author{Liam Doyle}
\author{Chandralekha Singh}
\affiliation{Department of Physics and Astronomy, University of Pittsburgh, Pittsburgh, PA 15260}

\begin{abstract}
We describe the development, validation and implementation of a Quantum Interactive Learning Tutorial (QuILT) on quantum key distribution (QKD) using entanglement, a context which involves a practical application of quantum concepts relevant for the second quantum revolution. The QuILT helps students learn quantum concepts relevant for quantum cryptography using a simple two-particle system. The protocol uses two entangled particles and two Stern-Gerlach Apparati to generate a random shared key over a public channel for encrypting and decrypting information. It actively engages students in the learning process and helps them build links between the concepts learned in class and their real world applications. The QuILT was implemented as a homework in a traditional quantum mechanics course and a quantum computing and quantum information course. The evaluation suggests that the QuILT is helpful in improving students' understanding of the concepts related to QKD in both courses. Also, the evaluation of the QuILT in both types of courses with or without lecture-based instruction in relevant QKD concepts suggests that the QuILT can be given as a homework after lecture-based instruction on entanglement without in-class discussion on QKD. Therefore, entanglement can be covered without taking up much in-class time, while also providing students with an understanding of the QKD method and how it protects from eavesdroppers.
\end{abstract}

\maketitle

\section{Introduction and Framework} \label{Intro}

Quantum information science and engineering (QISE) is facing a rapid growth due to the second quantum revolution \cite{european, raymer2019, flagship, divincenzo,liam}, which is bringing transformative advancements to the field. As QISE grows rapidly, the need for a more skilled and interdisciplinary workforce is increasing as well \cite{fox2020cu, meyer2022cu, muller2023prperworkforce, Kashyapmisinformation, ghimire2025reflections, fargol,liam2, hypeliam}. 
While considerations are being made about how universities can address this need, through university courses and curricula \cite{qtgoorney, singhasfaw2021pt, goorney2024framework}, new programs \cite{asfaw2022ieee, kohnle2013}, and even shorter courses which have the specific goal of interdisciplinary accessibility \cite{qthellstern}. However the difficulties that come with addressing these needs are compounded by the need for uniformity in the education that is being provided, with educators needing to identify underlying concepts which are most important to QISE \cite{qtmerzeletal, bungum2022quantum, donhauser2024empirical}.

Quantum concepts are challenging even for advanced undergraduate students, since they are often abstract or counterintuitive and the quantum paradigm is very different from the classical paradigm \cite{singh2015review, marshman2019qmfps}. Therefore, there has been much research on how to address these difficulties, with considerations of how to introduce these topics at the high school level \cite{hennig2024new, qtsun, weissmanphysics, rodriguez2020designing, ghimire2025epj, michelini2022, singh2022tpt, bondani}, as well how to address difficulties that often arise at the university level \cite{singh2007comp, qtmeyercu, jeremyajp, marshman2015}. At the university level, there has been  research about common topics and examples in quantum, like entanglement \cite{qtbrang}, quantum optics (commonly using the example of the Mach-Zehnder Interferometer) \cite{justicemathphysics, marshman2016ejpphoton, maries2020mzidouble, bitzenbauer3}, and Quantum Key Distribution (QKD) \cite{devore2020qkd, qtmerzel}. Often these topics are supplemented using interactivity, either through simulations and visualizations \cite{Kohnle_2017, Hubloch, schalkers2024explaining, michelini2023research, Benlarmorajp2025}, experiments which can be completed in an undergraduate laboratory \cite{beckphoton, kiko, nvcenter2}, or even physical tools which can assist with understanding in the classroom \cite{lopez2020encrypt, jeremytpt}.

We have been using research on student difficulties in learning quantum concepts as a guide to develop learning tools, e.g., a set of quantum interactive learning tutorials (QuILTs) that provides background on quantum information including basics of quantum computing, operators, probability distribution for measuring outcomes, quantum measurement and expectation values, quantum optics using single photons, e.g., see Refs. \cite{singh2007comp,zhu2012measure2, justicemathphysics, hucomputing, Hubloch, devore2020qkd, marshman2016ejpphoton, marshman2017expect, marshman2017prob, marshman2017opejp}. These concepts could be covered in quantum mechanics courses. 

Here, we discuss student learning from a QuILT related to quantum cryptography \cite{bennett1984,ekert1991quantum,kwiatqkd,RevModPhysqkd,qkdpanchina2017} on quantum key distribution (QKD) using two entangled spin-1/2 particles. QKD is an application of quantum concepts relevant for the second quantum revolution \cite{NSAqkdsecurity}. Although often confused with sending encrypted messages, QKD protocols are used to create a secret shared key between two parties that is then used to send an encrypted message over a public channel that will later be decrypted by the receiver using the same key. There are many different protocols for QKD, e.g., see \cite{bennett1984, ekert1991quantum, bennett1992quantum, B92protocol}. 

The QKD protocol allows two parties, a sender (who is usually referred to as Alice) and a receiver (who is usually referred to as Bob) to be able to generate a random shared key over a public channel (without meeting in person). The unique feature of QKD as opposed to classical approaches to key distribution is that it is secure in that Alice and Bob can detect the presence of an eavesdropper (who is usually referred to as Eve), who may be intercepting their communication to attempt to learn what the secret shared key between Alice and Bob is. Alice and Bob are connected via a channel involving an entangled quantum state (involving two spin-1/2 particles in the QuILT), which they use as a resource to generate a random shared secret key. More detail about this protocol will be provided below in Section \ref{protocol}.  QKD can be a helpful example for students to see how concepts they've learned can be useful in the quantum information revolution, e.g., for quantum cryptography.  Some QKD protocols are currently being explored for use, e.g., in the banking industry \cite{quantique}. 

Previously, learning tools on QKD including lab experiments \cite{galvezqkd}, simulations \cite{Kohnleqkd}, and a textbook \cite{woottersqkd} on cryptography (which has QKD) for college level have been developed. Learning tools and outreach programs at the pre-college level on QKD have also been developed \cite{qtmerzel,kellyqkd,netoqkdoutreach}. Also, we previously developed and validated a tutorial on QKD using non-orthogonal polarization states of single photons \cite{devore2020qkd}.

The tutorial developed and validated as part of the research described here relies on entanglement to generate a shared key between two remote parties. Our research shows that QKD using entanglement is an example which allows students to learn about how entanglement can be harnessed in a practical application. We found that students tended to engage very well with this protocol with little difficulty after learning conceptually about entanglement in class, usually in the form of a lecture which provides students with both a conceptual and mathematical understanding of entanglement. 
In particular, the lecture consisted of Bell states and how these states are not product states (not factorizable into products of two different entities) and measurement on one entity correlates with measurement outcome of the other entity.

Similar to our other QuILTs, we use Vygotsky's framework emphasizing the importance of providing students appropriate scaffolding support while remaining in their zone of proximal development (ZPD) \cite{vygotsky}. The ZPD is defined by what a student can do on their own without scaffolding vs. what they can do via instruction using learning tools, such as the QuILTs, that are designed to build appropriately on their prior knowledge, e.g., see \cite{dorisquantum2025year}. Furthermore, we take inspiration from the preparation for future learning (PFL) framework which suggests that effective instructional approaches should involve both innovation and efficiency dimensions to help students be in the optimal adaptability corridor and transfer their learning from one situation to another \cite{schwartzbransford,schwartz1998time}. In the context of the QuILT on QKD, students are asked to answer questions that represents the ``innovation" dimension and then scaffolding support is provided via the guided teaching learning sequences, which represents the ``efficiency" dimension. Through a variety of assessment questions that students must answer after engaging with the tutorial, we investigate transfer of learning in both near and far transfer contexts. For example, assessment tasks probing student understanding after engaging with the QKD tutorial include cases in which students are asked (near transfer) questions that are very similar to what they learned from the tutorial as well as those in which they are asked far transfer questions that are not related to concepts explicitly discussed in the tutorial. 

The final version of the tutorial is included in the supplemental materials and the pre-/post-test (which are the same) is included in the Appendix. As noted, another QuILT was developed and validated 
earlier that helps students learn QKD using two non-orthogonal polarization states of photons \cite{devore2020qkd} and has been found effective. 
The main difference between these two approaches to doing QKD is the physical mechanism with the one we discuss here for helping students learn involving use of entanglement as a resource instead of non-orthogonal polarization states of single photons. QKD takes advantage of the fact that quantum states, in general, cannot be cloned and measurement can irreversibly change the state \cite{nocloning}.

\subsection{The QKD Protocol students learn in this QuILT} \label{protocol}

\begin{figure}[h!]
    \centering
    \includegraphics[scale = 0.75]{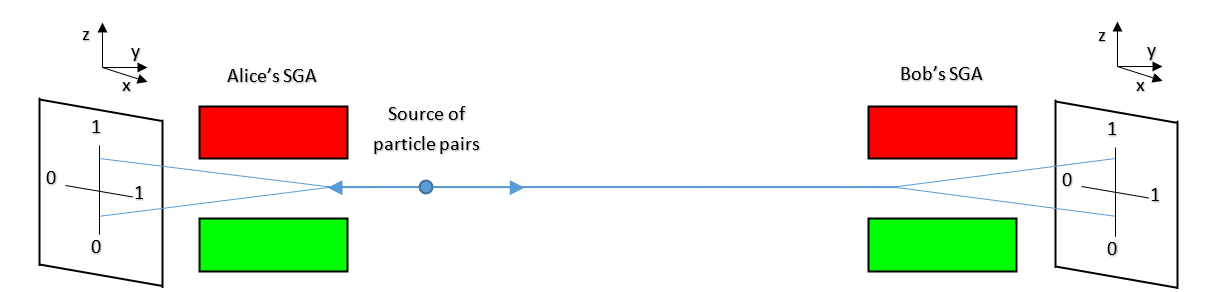}
    \caption{Depiction of Alice and Bob's setup along with reference axes.}
    \label{fig:Alice_Bob_setup}
\end{figure}

In the setup used in the QuILT, Alice and Bob each have a Stern-Gerlach Apparatus (SGA) and a screen, which acts as a measurement device (it flashes when a particle hits the screen). A schematic diagram of the setup is provided in Figure \ref{fig:Alice_Bob_setup}. The SGA consists of a region of non-uniform magnetic field, which has the field gradient along a given axis, e.g., X or Z. This inhomogeneous magnetic field interacts with the spin angular momentum of the particle passing through and causes deflection that aligns with the axis of the SGA (up/down or left/right depending upon whether the SGA has the magnetic field gradient aligned with the Z-axis or X-axis, respectively). The measurement of the spin of the particle along the X or Z axis can be made by using a screen to detect how the particle is deflected. 

In the tutorial, students are told that an entangled particle pair in the state $\ket{\Psi_{AB}} = \frac{1}{\sqrt{2}} (\ket{\uparrow_{A}} \ket{\downarrow_{B}} - \ket{\downarrow_{A}}\ket{\uparrow_{B}})$ is created, such that particle A in the entangled pair is in Alice's lab and particle B is in Bob's lab. All notations are standard and A and B subscripts are shorthands for Alice and Bob's particle's states in the entangled pair they use as a resource for QKD. They then decide on a convention, e.g., that up and right will be measurement outcome 1 and down and left will be measurement outcome 0, but since the entangled state is an anti-correlated one, when Bob measures a ``0'' he will record a ``1'' in their shared key and vice versa. After each measurement, over a public channel, they will state which basis (X or Z) each of them used for their measurement, but not what result they obtained so that they can determine which bits to keep in their shared key. They then continue this until they generate a secret shared key comparable to the length that they want, which will allow them to send their encrypted message over a public channel without anyone else having knowledge of the key they would use for encrypting and decrypting it. 

In the QKD approach in the QuILT, if there is an eavesdropper, she intercepts every particle which is sent to Bob with her own SGA that she randomly orients along the X or Z direction (similar to Alice and Bob's strategy). After intercepting a particle, Eve immediately generates a replacement particle to send to Bob in its place, sending the particle in the basis state that she measured in (e.g., if she measures a 1 in the X basis, she will send to Bob that state). However, Alice and Bob's protocol will allow them to detect Eve's presence. Every time Eve measures, there's a 50\% probability that Alice and Bob will select the same basis (therefore they keep the bit). Eve has a 50\% chance that she chooses the same basis as Alice and Bob and remains undetected. However, if Eve is in the opposite basis, she has another 50\% chance of remaining undetected as Bob's measurement could still lead to an output that's anti-correlated with Alice's. This means that 75\% of the time Eve is undetected, but 25\% of the time Alice and Bob will keep an outcome as part of their shared key generated that is supposed to be discarded. Therefore, if they compare enough bits over a public channel (e.g., by doing a parity check, and do not include these bits in their shared key), they will find that in 25\% of the cases, their shared key has bits that should not be there (which is more than the noise level) and they will know that there is an eavesdropper. They can then discard the key and stop the process of generating their shared key and begin again another time when there is no eavesdropper present. The tutorial also integrates a QuVis simulation \cite{Kohnleqkd} on QKD using entanglement to solidify these concepts.

\subsection{Research Objectives and Questions}

The primary objective of this research is to develop and validate a QuILT about QKD utilizing entanglement and evaluate its efficacy in helping students learn relevant concepts. After the development and validation, students in two different types of classes engaged with the QuILT, with one class being a typical undergraduate quantum mechanics course, and the other class being an interdisciplinary foundations of quantum computing and quantum information course for any science and engineering undergraduate major, who had taken two calculus courses with a $C^+$ grade (this math requirement is to ensure that students have sufficient mathematical reasoning skills so that they can learn any linear algebra that the instructor considers necessary in the course). An important goal of the QuILT is to provide appropriate scaffolding support while being in students' ZPD and be effective for students from both physics and interdisciplinary backgrounds pertaining to both near and far transfer assessment questions. Thus, we wanted to ensure that the tutorial is accessible to students in both types of courses (i.e., a course exclusively for physics majors and an interdisciplinary course for students from any science and engineering majors).
Specifically, for students in these two types of courses, we address the following research questions:

\begin{itemize}
    \item[\textbf{RQ1.}] How does student performance on assessment questions change between traditional lecture-based in-class instruction on QKD using entanglement and after engaging with the QuILT?

    \item[\textbf{RQ2.}] How do students perform on assessment questions after engaging with the QuILT if there is no in-class instruction on QKD using entanglement and students engage with the QuILT after in-class instruction on entanglement only?

    \item[\textbf{RQ3.}] Are students in a quantum mechanics course able to perform comparably on assessment questions after engaging with the QuILT regardless of whether they received in-class instruction on QKD?

    \item[\textbf{RQ4.}] Are students in an interdisciplinary course able to perform comparably on assessment questions  after engaging with the QuILT regardless of whether they received in-class instruction on QKD?
\end{itemize}

\vspace*{-0.2in}
\section{Methodology} \label{Methodology}
\vspace*{-0.1in}

\subsection{Participants and Course Details} \label{participants}

The participants from the in-class implementation of the QuILT are four types of groupings of five separate undergraduate classes (labeled Class 1 through Class 4 with sub-labels to denote the type of course) at a large public research university in the US. The groups are separated by course and by the treatment they had with respect to either learning QKD (using entanglement) in class or only learning about entanglement in class before engaging with the tutorial (discussed in further detail below).

Class 1-QM and Class 4-QM both consist of students from a quantum mechanics (QM) course, which is intended for students majoring in physics. This course is taken mostly by undergraduate students who are in their 3rd or 4th year. Class 1-QM consists of one class of students in this QM course. The course instructor for Class 1-QM utilized the textbook \textit{Introduction to Quantum Mechanics} by David Griffiths \cite{griffithsQM} to guide the course including for homework problems.
Class 4-QM consists of students from two different sessions (in two different semesters) of this same QM course. The course instructors for Class 4-QM for these two consecutive years were different but they both utilized the textbook \textit{A Modern Approach to Quantum Mechanics} by John S. Townsend \cite{townsend2000modern} to guide the course and for homework problems. These two course instructors for Class 4-QM were also different from the course instructor of Class 1-QM. Class 2-QCQI and Class 3-QCQI are from different years of a foundations of quantum computing and quantum information (QCQI) course which is intended for any undergraduate students majoring in computer science, engineering, chemistry, mathematics, physics etc. (who have obtained at least a $C^+$ grade in two introductory calculus courses) and does not have a requirement for any prior exposure to quantum mechanics (e.g., via a modern physics course). The course is therefore accessible to a broader audience; it has been taken by students in any of the years (first-year through fourth-year) of undergraduate study and of many different majors. For Class 2-QCQI,  the course instructor did not use specific textbooks, instead using a variety of resources to develop a course which can provide students with information about quantum computing and quantum information. For Class 3-QCQI, the course instructor utilized the textbook \textit{Introduction to Classical and Quantum Computing} by Thomas G. Wong \cite{wong2022introduction} to guide the course and for homework. Class 1-QM had N = 22 students, Class 2-QCQI had N = 11, Class 3-QCQI had N = 12, and Class 4-QM had N = 31 students.

The students in Class 1-QM and Class 2-QCQI first learned QKD via an in-class lecture, and therefore were given a pre-test to assess the efficacy of the traditional lecture-based instruction on QKD. Their pre-test performance could then be compared with the post-test performance after engaging with the tutorial, which was given as a homework. 
The students in Class 3-QCQI and Class 4-QM were given no in class instruction on QKD, and were instead just given the QuILT as a homework after learning about entanglement via lecture and then given the post-test in class. The information about these classes is summarized in Table \ref{tab:classinfo}.

In summary, the pre-test and post-test (which are identical) were always administered in class and the QuILT was always administered as homework in all four types of classes. The QKD QuILT was graded as a homework for each course. However, only Class 1-QM and Class 2-QCQI had an in-class lecture on QKD so only they were given a pre-test. Class 3-QCQI and Class 4-QM took only the post-test after the QKD QuILT since they only had traditional lecture on entanglement related to Bell States and how these entangled states are not product states and measurement outcomes of two entities are correlated (but no traditional lecture on QKD).

\begin{table}[h!]
    \centering
    \begin{tabular}{|c|c|c|c|c|c|c|}
        \hline
         Course & Class Name & Number of Students & In-Class Instruction  & Pre-test Given & Tutorial as Homework & In-class Post-test\\
         \hline
         QM & 1-QM & 22 & Yes & Yes & Yes & Yes\\
         QCQI & 2-QCQI & 11 & Yes & Yes & Yes & Yes\\
         QCQI& 3-QCQI & 12 & No & No & Yes & Yes\\
         QM & 4-QM & 31 & No & No & Yes & Yes\\
         \hline
    \end{tabular}
    \caption{Information about the four types of classes that participated in this study. Information regarding the course including the name associated with the given class type, the number of students in the class, whether or not they received in-class lecture-based instruction on QKD, whether the class was administered a pre-test, whether they engaged with the tutorial as a homework, and whether they completed the post-test in their class is provided.}
    \label{tab:classinfo}
\end{table}

\subsection{QuILT Development and Validation} \label{dev_and_val}

Before we describe the learning objectives, we note that there are four Bell states, which are maximally entangled states involving two-qubits. The entangled state that Alice and Bob use in the QKD protocol described earlier is one of them. All four Bell states are provided in the Appendix as part of the assessment question 4 on the pre-test and post-test. 

The learning objectives (LO) of the QKD QuILT with entanglement are as follows:

\begin{itemize}
    \item[\textbf{LO1.}] Students are able to describe the protocol that Alice and Bob have agreed upon for generating the shared secret key for encryption.

    \item[\textbf{LO2.}] Students are able to identify and explain how Alice and Bob's protocol guarantees the detection of Eve's presence.

    \item[\textbf{LO3.}] Students are able to identify and explain why other maximally entangled Bell states (other than the Bell state used in the protocol students learned) would also be appropriate for the QKD protocol involving entanglement but product states would not be appropriate.
\end{itemize}

We note that the learning objectives are also organized in order of transfer distance from the point of view of research on transfer of learning \cite{schwartzbransford,maries2020mzidouble}, where LO1 is a near transfer learning objective (since the assessment question is identical to what students learned in the protocol provided in the tutorial and assessment), LO2  requires intermediate-level of transfer since students must synthesize what they learned from the QuILT about how the presence of an eavesdropper will be detected and transfer it in verbal representation \cite{cedricrepresentation2012,representation2024,representationlin2017}, and LO3 is a far transfer question since the QuILT does not provide specific scaffolding relating to LO3 (although all students had separately learned about the four Bell states not in the context of QKD).

There are four questions on the assessment used for pre-/post-testing given in the Appendix (pre-/post-tests are identical). We note that Question 1 (Q1) and Q2 are identical to two questions in the tutorial and the protocol provided for QKD using entanglement in pre-/post-test. 
This is because we wanted to assess that students are able to, at the very least, explain how the QKD protocol between Alice and Bob works without the presence of an eavesdropper. Q3 on the assessment focuses on the role of the eavesdropper and students are asked to apply what they have learned to explain in verbal representation (words) why the presence of the eavesdropper will be detected. The last question on the assessment, Q4, is a far transfer question. Q4 has several two-particle states (some of which are Bell states and others are product states) and students are asked to identity and explain which of those states would be appropriate to use for QKD with entanglement (even though students were not provided explicit scaffolding for this type of question).

Similar to the development and validation of the other QuILTs, e.g., 
on quantum key distribution using non-orthogonal polarization states of light \cite{devore2020qkd}, the development and validation of this QuILT (alongside the pre-test and post-test) began by assessing student difficulties after lecture-based instruction and using the research as a guide to provide scaffolding as needed in the QuILT. In particular, the QuILT builds on students' prior knowledge, determined via the investigation of difficulties as well as what students are able to accomplish with a given level of scaffolding support. It is reliant on an inquiry-based approach to build on students' knowledge while keeping them engaged in the learning process throughout the tutorial via the teaching-learning sequence \cite{zuza2023teachinglearning}. This inquiry-based guided sequence is based upon a cognitive task analysis \cite{taskkirwan} from both faculty and student perspective and is important as each question builds on the next, with considerations for both how to help students learn a concept and how to account for common student difficulties. Usually common student difficulties are addressed via presentation of a guided simulated conversation in which the student is presented with a common difficulty and they must identify why a given alternative conception is incorrect. The tutorial also provides students with a simulation \cite{Kohnleqkd} and asks them to utilize the simulation to check their predictions to prior questions. In an instance where a student's predictions do not align, they are asked to reconcile these differences.

The development and validation of the QuILT follows a cyclic process as follows:

\begin{enumerate}
    \item Development of a preliminary version based on a cognitive task analysis \cite{taskkirwan} of the underlying knowledge from both faculty and student perspectives including research on student difficulties with relevant concepts.
    \item Implementation and evaluation of the QuILT by administering it individually to students and receiving feedback from faculty members who are experts in these topics.
    \item Determining its impact on student learning and assessing what difficulties were not fully addressed by the QuILT.
    \item Refining and adjusting based on the feedback provided after the implementation and evaluation.
\end{enumerate}

In the most recent iteration of this process, we conducted think-aloud interviews with 9 advanced physics students via Zoom. The students who were interviewed were either from an undergraduate course on modern physics, which includes concepts relating to quantum mechanics, or were graduate students. Out of all these students, only 2 had previously engaged with QKD in their course before the interviews. The interviews were approximately an hour and a half long and allowed us to understand students' thought processes as they engaged with the QuILT and corresponding assessments regarding where further scaffolding was required, and if there was a need to finetune the questions so that they provide appropriate guidance to students. 

During these interviews, students were asked to think aloud as to how and why they answered questions as they did. If they had difficulties and could not proceed, they were given a brief hint; this usually indicated that further scaffolding may be required. 

We also had these students engage with another QuILT in individual interview settings that focus on helping them learn about QKD using non-orthogonal polarization states of light \cite{devore2020qkd} to compare the relative difficulty and effectiveness of the two different QKD approaches for students. Our findings on this comparison are discussed further in Section \ref{Performance}. 

Also, as noted, we iterated the QuILT with faculty members who are experts in the field and refined it by making adjustments based on their suggestions.

\subsection{In-Class Implementation}

The current version of the QuILT was also given to 5 separate classes (but comprising only four separate types of classes including whether students had in class instruction on QKD) as described earlier in Section \ref{participants}. 
All of these classes utilized the tutorial as a homework assignment, which was given to students with the understanding that they would be tested on their understanding of QKD via the post-test associated with the QuILT.
As noted in Section \ref{participants}, Class 1-QM and Class 2-QCQI were given in-class instruction on QKD and therefore were given a pre-test to assess their understanding of QKD. However, since Class 3-QCQI and Class 4-QM did not have in-class instruction on QKD (and had only learned about entanglement via lecture-based instruction), they were only given the post-test as they had no prior knowledge to compare to after they engaged with the QuILT. 
The post-test was given either on its own as a quiz after students had engaged with the QuILT or integrated into an exam which the students took soon after engaging with the QuILT. 

We note that students were not given direct feedback on the pre-test nor the QuILT itself. Instead, students in these courses were simply given credit for completion on pre-test as to not sway the data of the post-test by providing them with feedback.

\begin{figure}
    \centering
    \includegraphics[width=1.0\linewidth]{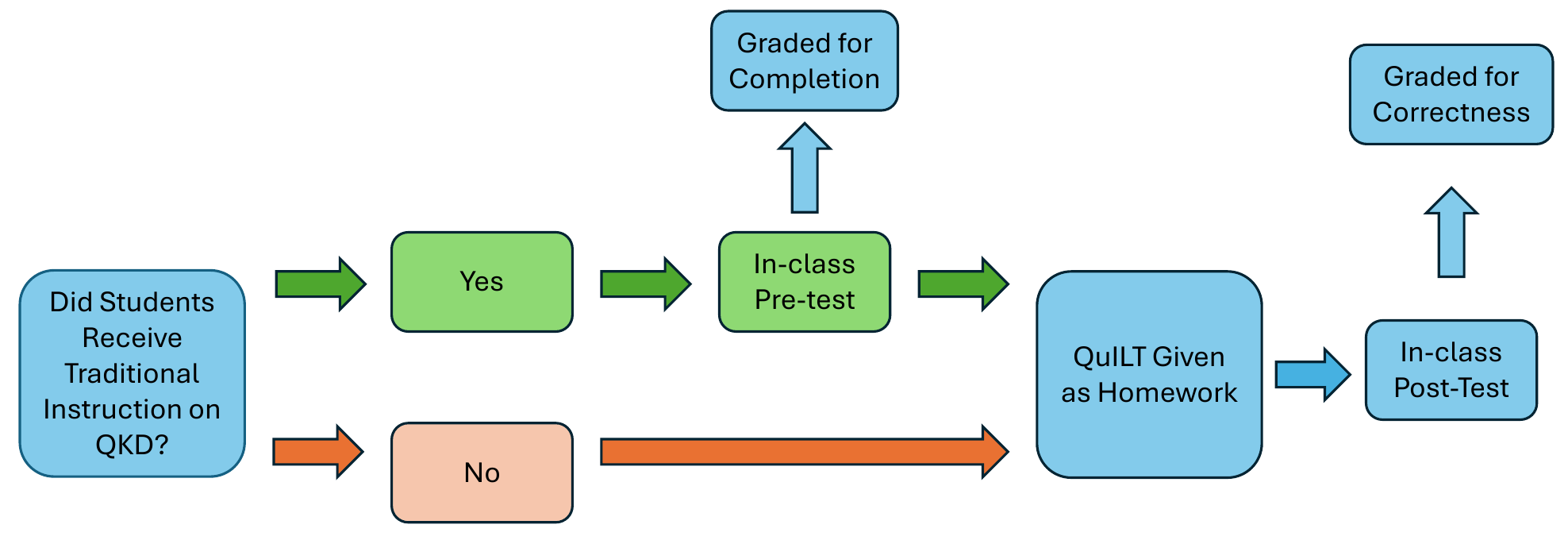}
    \caption{Visual representation of how the QuILT (along with accompanying assessments) was implemented for all classes.}
    \label{fig:Implementation}
\end{figure}

A visual representation of how the QuILT and the corresponding assessments were implemented is represented schematically in Figure \ref{fig:Implementation}.

\subsection{Data Analysis} \label{DataAnalysis}

After the in-class implementation of the QuILT, we scored the pre-test and post-test to investigate student learning based upon their performance on the assessments in the different classes. The rubric used to score the responses to the pre-test and post-test is summarized in Table \ref{table:rubric} and was developed collaboratively by researchers.

\begin{table}[b]
    \centering

    \begin{tabular}{|c|p{1.82in}|p{0.8in}|p{2.2in}|}
    \hline
         Question Number & Learning Objective Covered & Max Score & Explanation of Point Breakdown  \\
         \hline
         1 \& 2 & LO1: Describe the protocol Alice and Bob have agreed upon for creating a shared key for encryption. & Q1: 4 points Q2: 8 points & -1 point for each incorrect response. \\
         \hline
        3 & LO2: Describe how the eavesdropper's presence would be detected & 2 points & -1 point for not identifying that Eve could only be detected when she is in a different basis than Alice and Bob. -1 point for not identifying Eve's detection is probabilistic. \\
        \hline
        4 & LO3: Identify which states, other than the one provided in the tutorial, are suitable to use with this QKD method and explain the choices. & 6 points & -1 point for each correct option not identified (of which there are 3). -1 point for selecting each incorrect option (of which there are 2). -1 point for not explaining why the other Bell States (options A, B, and D, as seen in the Appendix) are suitable for QKD.\\
        \hline
    \end{tabular}
    \caption{Table shows the rubric used for the pre-/post-test and provides explanations for what students need to be able to describe/identify/explain, how many points each question is worth, and an explanation of how the points were deducted.}
    \label{table:rubric}
\end{table}

Despite the initial different point distributions for each question to make it easier (questions were distributed by parts), in the end, the score of each question is normalized so that each question is worth the same amount. 
Thus, each question is weighed equally to the others when assigning an overall score for students' performance on the pre-/post-test.

For Question 1 (Q1) and Q2, no interrater agreement was required as each response was only correct or incorrect. For Q3, researchers separately scored 50\% of the responses and found that their scores for all of these responses had perfect interrater agreement. Similarly, for Q4, the researchers separately scored 50\%  of the responses and found again that their scores for all of these responses had perfect interrater agreement.

To compare the results of pre-tests to post-tests within Class 1-QM and Class 2-QCQI, we calculated effect size using Cohen's ${\it d}$ 
\cite{cohen} 
as follows: 

\begin{equation}
    d = \frac{\overline{x_{post}} - \overline{x_{pre}}}{s}
\end{equation}

\noindent in which $\overline{x}$ represents the mean for a given class and $s$ is the pooled standard deviation which is given as follows:

\begin{equation}
    s = \sqrt{\frac{(n_{post}-1)s_{post}^{2} + (n_{pre}-1)s_{pre}^{2}}{n_{post}+n_{pre}-2}}
\end{equation}
in which $s_{pre}$ and $s_{post}$ are the standard deviations of the pre-test and post-test, respectively, and $n_{pre}$ and $n_{post}$ are the number of students who took the pre-test and post-test, respectively.

To check for statistically significant differences between pre-test and post-test, the ${\it p}$-value was calculated as well using a paired t-test using the pre-test and post-test data for each question. The p-values for each question are provided for both Class 1-QM and Class 2-QCQI. 

We note that on some questions there is no statistically significant change in performance between pre-test and post-test, therefore the Cohen's ${\it d}$ is not provided.

\vspace*{-0.2in}
\section{Student Difficulties and How the Tutorial Addressed Them} \label{Difficulties}
\vspace*{-0.1in}

We find that after traditional lecture-based instruction on QKD, students struggled with the security of the QKD protocol and how the presence of an eavesdropper would be detected when utilizing QKD protocol (with entanglement) to generate a shared secure key. This difficulty is assessed by asking students to identify why an eavesdropper's presence would be detected when using the given protocol (see pre/post-test Q3 in the Appendix). After lecture-based instruction, one student wrote, ``Eve would be detected because once she measures, the entangled state of the qubit will collapse with her measurement. If Eve sends a replacement bit, it will be known because that bit will not be entangled with Alice's bit''. This student correctly states that Eve's measurement will collapse the entangled state. However, the student incorrectly claims that the replacement bit not being entangled with Alice's bit can be taken as the reason that Eve's presence would be detected. 
This notion is one that's seen sometimes in the pre-test, in which a student 
states that just because the receiver isn't sent an entangled bit that they would know there's an eavesdropper, rather than needing to do, e.g., a (parity) check at the end to determine if a fraction ($\sim$ 25\%) of the results are incorrect and therefore know that an eavesdropper is present (when Alice and Bob's SGAs are aligned but Eve's SGA is not aligned with them). To address this difficulty, when students engage with the tutorial about the interference of the eavesdropper, they are provided support to identify whether or not the eavesdropper's presence would be detected for given orientations of their SGAs. The tutorial guides students  
to help them recognize the effect that an eavesdropper has and how the security of this QKD method is guaranteed by the approximately 25\% error rate being introduced by the eavesdropper.

Another difficulty that repeatedly occurred during interviews, but was not tested in pre and post-test, was on questions in the tutorial in which students are asked to identify the percentage certainty of Bob measuring a `1' if Alice measures either a `0' or `1' when there is no eavesdropper present. Often students expressed that they were equally conflicted between choosing 50\% or 100\%. For example, one such student stated ``It's either [100\% certainty] or [50\% certainty]...if [Alice and Bob are] not oriented correctly, then it doesn't mean they're guaranteed to match up. So that means [Bob] would still have a 50/50 chance of [measuring a 1], but if that's not a factor [they are both using the same basis], then [Bob measuring a 1] would be 100\%''. This student then selected the option of 50\%. Interviews suggest that this type of difficulty may be due to the fact that students did not consider using conditional probabilities to determine what would happen when Bob chose the X or the Z basis. In particular, this student correctly identified the individual probabilities for when Bob is in each given state. However, they did not utilize conditional probability to determine that the overall likelihood is 75\%, and therefore only ended up stating that there was a 50\% chance of Bob measuring a 1 when Alice measures a 0.
While this difficulty is not necessarily related directly to QKD, we found it important to address because if the difficulty remained, the issue would reappear later when students are asked about the probability of an eavesdropper to be detected. This difficulty was addressed in the tutorial with the addition of part b to question 11, which provides scaffolding support via the correct probabilities and asks students to write out the probability calculation if their values do not align with the correct values. 
This was helpful guidance because in interviews, students who were unable to obtain the correct probabilities, but were also prompted by the interviewer to write out their approach to calculations, noted that the action of writing out the probability assisted them in answering this question correctly. 
These students were then able to successfully identify the probability of an eavesdropper's presence being detected later in the tutorial. 

In response to Q4 on the pre-/post-test, we often found that students who struggled would mostly select only the Bell state in choice B of Q4 as appropriate for QKD, which is similar to the singlet entangled state in the tutorial (anti-correlated) but with a positive sign (see option B of Q4 in the Appendix). For example, after lecture-based instruction, one student who chose only choice B wrote, ``You need to know the opposite state from your measurement [for QKD]". Similarly, even after engaging with the tutorial, one student responded ``[Choice B is] the only other state where they are properly anti-correlated" suggesting the state for QKD could not be the other two Bell states (see options A and D in Q4 in the Appendix). This appears to imply that this student had an alternative conception that the anti-correlation in the entangled state in the tutorial is a key feature in the QKD, not that the important feature for QKD is the maximally entangled state, e.g., represented by any of the four Bell states. Due to this difficulty persisting in the post-test responses to Q4, we added two questions to the tutorial to improve student understanding. The first question that was added is a discussion between three hypothetical students that focuses on different Bell states (maximally entangled states) that can be used for QKD, with one student saying that all Bell states are good choices, another student saying only the given state in the tutorial is a good choice, and a third student saying only the two anti-correlated Bell states (with plus or minus signs) would be good choices. The second question added in the tutorial gives students a product state (a state that is separable) and asks students to identify if it is possible to do QKD using this state and to explain why or why not. These discussions in the tutorial are intended to guide students to think deeply about the reasoning behind what entangled states could work for QKD. Students who engaged with the tutorial during an interview were able to use these as a means to help them with the synthesis that Q4 is hoping to achieve, though we have yet to give that version of the tutorial to a class of students as an integral part of the course.

Interestingly, in the written responses on the pre-/post-test, we found that students appeared to have a difficulty with the idea of correlation. In some responses, students emphasized the importance of ``correlation" between the results, but usually did so in the context of if one result is up then the other must be down. We note that this is not totally accurate as, by this definition, a state like $\ket{\uparrow_{A}\downarrow_{B}}$ could be considered anti-correlated when this is not the case, since the measurement of the `A' particle will have no 
correlation with the `B' particle's measurement (there should at least be two possible outcomes for the `A' and `B' particles for their measurement outcomes to be correlated with each other). Although not dealt with directly in this tutorial, we found it important to note this point as we saw ideas related to this in some of the responses.

We note that during our interviews, we also had students engage with a different tutorial and associated assessments that helps students learn QKD using non-orthogonal polarization states of single photons \cite{devore2020qkd}. We found that the overall level of difficulties with QKD using non-orthogonal polarization states of photons 
were comparable to those on QKD using entanglement. 
However, we find that interviewed students thought that the QKD with entanglement was an effective means of learning about a physical application of entanglement. Therefore, they found this tutorial to be particularly helpful in this regard.

\vspace*{-0.2in}
\section{Results and Discussion} \label{Results&Discussion}
\vspace*{-0.1in}

\subsection{Performance on Pre-/Post-tests of the four Groups} \label{Performance}

The researchers discussed and decided that it is most appropriate to break down the performance comparisons into a few different categories: Classes that received in-class instruction on QKD (Class 1-QM and Class 2-QCQI), Classes that did not receive in-class instruction on QKD (Class 3-QCQI and Class 4-QM), QM Classes (Class 1-QM and Class 4-QM), and QCQI Classes (Class 2-QCQI and Class 3-QCQI)

Observation of the post-test scores across all four types of classes in Table \ref{table:Class1vClass2} and Table \ref{table:Class3vClass4} shows that all of the classes performed well after the tutorial on most questions. Figure \ref{fig:pre_post_comp} shows the comparisons of performance between the pre-/post-test for Class 1-QM and Class 2-QCQI. The detailed performance breakdowns for each question are provided in Figure \ref{fig:all_Grade_Breakdowns}, which were determined using the rubric shown in Table \ref{table:rubric}.

\subsubsection{Classes That Received In-Class Instructions on QKD}

\begin{table}[ht]
    \centering
    \begin{tabular}{|c|c c c c |c c c c |}
    \hline
       \multirow{2}{*}{\textbf{Question Number}} &
      \multicolumn{4}{c|}{\textbf{Class 1-QM}} &
      \multicolumn{4}{c|}{\textbf{Class 2-QCQI}} \\

    & Pre & Post & Effect Size & {\it p}-value & Pre & Post & Effect Size & {\it p}-value\\
    \hline
    1 & 98\% & 100\% & - & 0.164 & 86\% & 100\% & - & 0.096\\
    2 & 94\% & 100\% & 0.53 & 0.048 & 77\% & 100\% & 0.93 & 0.027\\
    3 & 66\% & 91\%  & 0.81 & 0.004 & 45\% & 82\%  & 0.89 & 0.016\\
    4 & 80\% & 86\%  & - & 0.182 & 79\% & 86\%  & - & 0.241\\
    \hline
    Average & 85\% & 94\% & - & - & 72\% & 92\% & - & -\\
    \hline
    \end{tabular}
    \caption{Pre- and post-test performance comparison for Class 1-QM and Class 2-QCQI, which received in-class instruction on QKD. Effect sizes given by Cohen's {\it d} is provided and {\it p}-values to show statistical significance are included. Only pre-/post-test differences on Q2 and Q3 are statistically significant according to the p-value ($p<0.05$), partly due to small $N$, therefore the Cohen's {\it d} is provided for those questions.}
    \label{table:Class1vClass2}
\end{table}

\begin{figure}
    \centering
    \includegraphics[scale = 0.6]{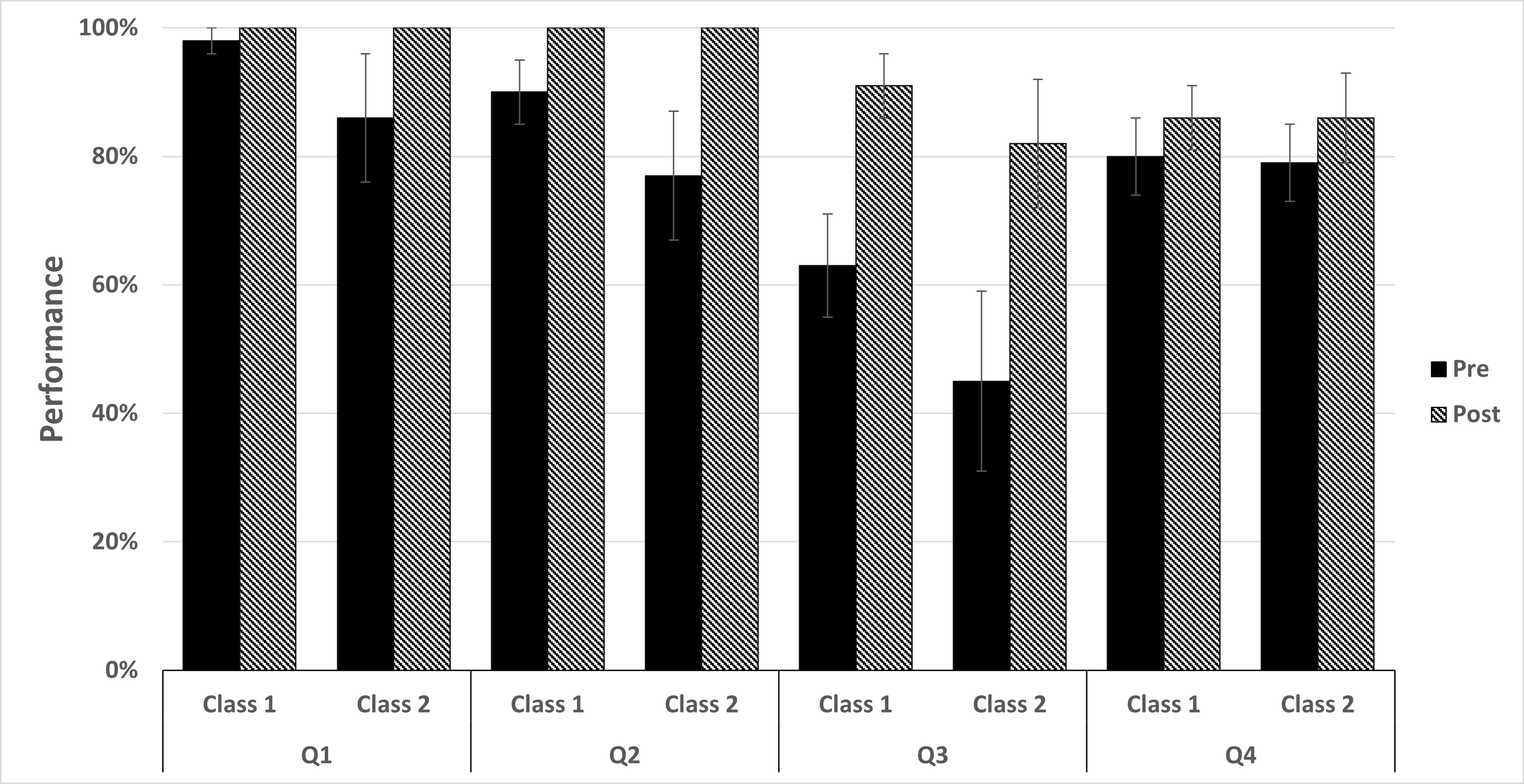}
    \caption{Bar chart representing the pre-test and post-test performances from Class 1-QM and Class 2-QCQI, as reported in Table \ref{table:Class1vClass2}. The bars are grouped by question number and class. The error bars represent the standard error.}
    \label{fig:pre_post_comp}
\end{figure}

The results of the pre-test and post-test for Class 1-QM and Class 2-QCQI are shown in both Figure \ref{fig:pre_post_comp} and Table \ref{table:Class1vClass2}.

We found that students did reasonably well on Q1 and Q2 of the pre-test, which suggests that students had a good grasp of how the QKD protocol works between Alice and Bob after lecture-based instruction (this information was also provided with the pre-/post-test as a description of the protocol used). However, for Q3 and Q4, there is some improvement from pre to post-test (although not statistically significant on Q4). This improvement suggests that, for these questions, the tutorial provided students with further support to assist them in developing a deeper understanding of the underlying concepts of QKD. For Q3, there was a large effect size ($d>0.8$) for both Class 1-QM and Class 2-QCQI from the pre-test to the post-test. This suggests that the tutorial was helpful in assisting students with how QKD is a secure means of key distribution and students were able to correctly describe why the introduction of an eavesdropper would not go unnoticed. For Q4, there was a small improvement for both Class 1-QM and Class 2-QCQI from the pre-test to the post-test, though other Bell states that could be used for QKD were not something that students specifically worked on in the tutorial version students in the four classes used (although they had all learned about Bell states separately as maximally entangled states via lecture-based instruction before engaging with the tutorial).

\vspace{-0.19in}
\subsubsection{Classes That Did Not Receive In-Class Instruction on QKD}

\begin{table}[ht]
    \centering
    \begin{tabular}{|c|c|c|}
    \hline
       \multirow{2}{*}{\textbf{Question Number}} &
      \multicolumn{1}{c|}{\textbf{Class 3-QCQI}} &
      \multicolumn{1}{c|}{\textbf{Class 4-QM}}\\
     & Post & Post \\
    \hline
    1 & 96\% & 97\% \\
    2 & 88\% & 96\% \\
    3 & 88\% & 85\% \\
    4 & 96\% & 72\% \\   
    \hline
    Average & 92\% & 88\% \\
    \hline
    \end{tabular}
    \caption{Post-test performance of Class 3-QCQI and Class 4-QM, which did not receive in-class instruction on QKD, instead only engaging with the topic through the QuILT as a homework assignment after learning about entanglement.}
    \label{table:Class3vClass4}
\end{table}

The results of the post-test for Class 3-QCQI and Class 4-QM are shown in Table \ref{table:Class3vClass4}. These findings suggest that students did reasonably well on Q1 and Q2 after engaging with the QuILT. For Q3, both classes did well which suggests that, without in-class instruction and only engaging with the QuILT as a homework, the students are able to understand how Eve's presence disrupts the process. For Q4, we found that students still did well especially when considering that, as previously mentioned, they engaged with the version of the tutorial which did not include scaffolding relating to which states could be used for QKD. We find these results encouraging as these students only engaged with the QuILT as a homework, meaning that the instructors in these classes did not have to take any time from their usual lectures to teach about QKD.

\subsubsection{QM Classes}

\begin{table}[ht]
    \centering
    \begin{tabular}{|c|c c c c |c|}
    \hline

       \multirow{2}{*}{\textbf{Question Number}} &
      \multicolumn{4}{c|}{\textbf{Class 1-QM}} &
      \multicolumn{1}{c|}{\textbf{Class 4-QM}}\\
    & Pre & Post & Effect Size & {\it p}-value & Post \\
    \hline
    1 & 98\% & 100\% & - & 0.164 & 97\% \\
    2 & 94\% & 100\% & 0.53 & 0.048 & 96\% \\
    3 & 66\% & 91\%  & 0.81 & 0.004 & 85\% \\
    4 & 80\% & 86\%  & - & 0.182 & 72\% \\   
    \hline
    Average & 85\% & 94\% & - & - & 88\%\\
    \hline
    \end{tabular}
    \caption{Performance comparison between QM classes.}
    \label{table:QM_Comparison}
\end{table}

When comparing the QM courses, as seen in Table \ref{table:QM_Comparison}, we find that Class 1-QM's pre-test score and Class 4-QM's post-test scores are very similar for both  Q1 and Q2. This shows that both classes had a good understanding of how Alice and Bob's strategy works to create their shared key after initial instruction (regardless of whether students had lecture-based instruction or engaged with the QKD tutorial without lecture-based instruction). However, on Q3, we see a more distinguishable difference in the understanding of why QKD is a secure protocol for students in Class 4-QM, who first engaged with the topic through the tutorial, in comparison to Class 1-QM, who first engaged with the topic through traditional lecture. When comparing the post-test results between Class 1-QM and Class 4-QM on Q3, it can be seen that this difference is eliminated, and that both classes did comparably on Q3. On Q4, we find that Class 1-QM did better than Class 4-QM after initial lecture-based instruction. This may be due to differences in the emphasis on Bell states in these classes before students engaged with the tutorial (although all instructors had discussed Bell states as maximally entangled states in their lecture before students engaged with the tutorial and corresponding assessments, their level of emphasis may have been different). Interestingly, we found that in Class 4-QM, 45\% of students scored 50\% on Q4, and all of those students had chosen B only as their answer. This response is something that was also persistent in the interviews, but Q4 was not a topic that was specifically addressed in the tutorial. Instead, as noted, it is a far transfer question that was included to investigate if students can synthesize what they've learned from the Bell state used in the tutorial for QKD using entanglement to the other Bell states (students had learned about Bell states separately as maximally entangled states in all courses).

\subsubsection{QCQI Classes}

\begin{table}[ht]
    \centering
    \begin{tabular}{|c|c c c c|c|}
    \hline

       \multirow{2}{*}{\textbf{Question Number}} &

      \multicolumn{4}{c|}{\textbf{Class 2-QCQI}} &
      \multicolumn{1}{c|}{\textbf{Class 3-QCQI}} \\
    & Pre & Post & Effect Size & {\it p}-value & Post \\
    \hline
    1 & 86\% & 100\% & - & 0.096 & 96\%\\
    2 & 77\% & 100\% & 0.93 & 0.027 & 88\%\\
    3 & 45\% & 82\%  & 0.89 & 0.019 & 88\%\\
    4 & 79\% & 86\%  & - & 0.241 & 96\%\\   
    \hline
    Average & 72\% & 94\% & - & - & 92\% \\
    \hline
    \end{tabular}
    \caption{Performance comparison between QCQI classes.}
    \label{table:QCQI_Comparison}
\end{table}

\begin{figure*}
    \centering
    \begin{subfigure}[b]{0.475\textwidth}
        \centering
        \includegraphics[width=\textwidth]{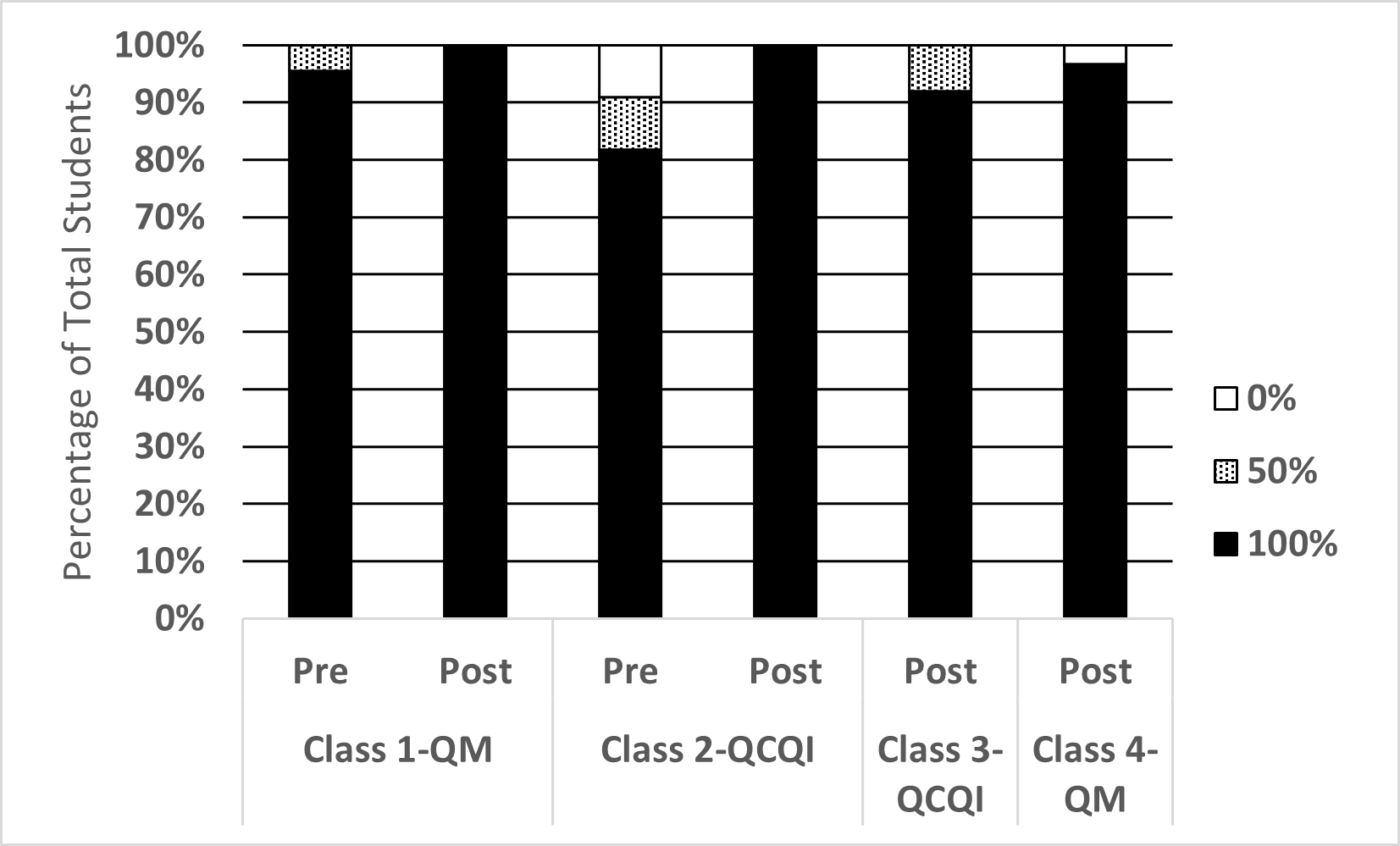}
        \caption[Q1]%
        {{\small Score breakdown for question 1.}}    
        \label{fig:Q1_Breakdown}
    \end{subfigure}
    \hfill
    \begin{subfigure}[b]{0.475\textwidth}  
        \centering 
        \includegraphics[width=\textwidth]{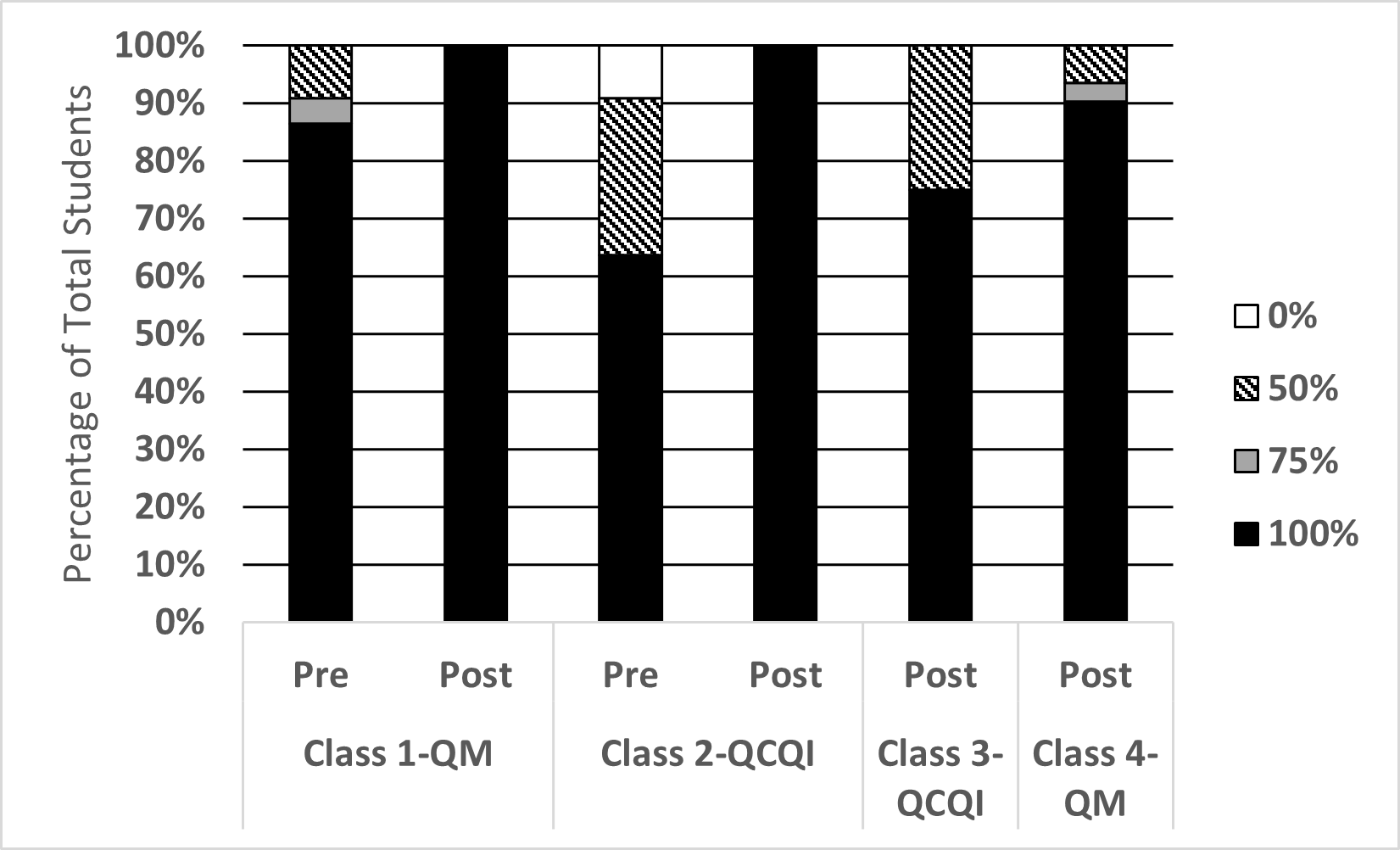}
        \caption[Q2]%
        {{\small Score breakdown for question 2.}}  
        \label{fig:Q2_Breakdown}
    \end{subfigure}
    \vskip\baselineskip
    \begin{subfigure}[b]{0.475\textwidth}   
        \centering 
        \includegraphics[width=\textwidth]{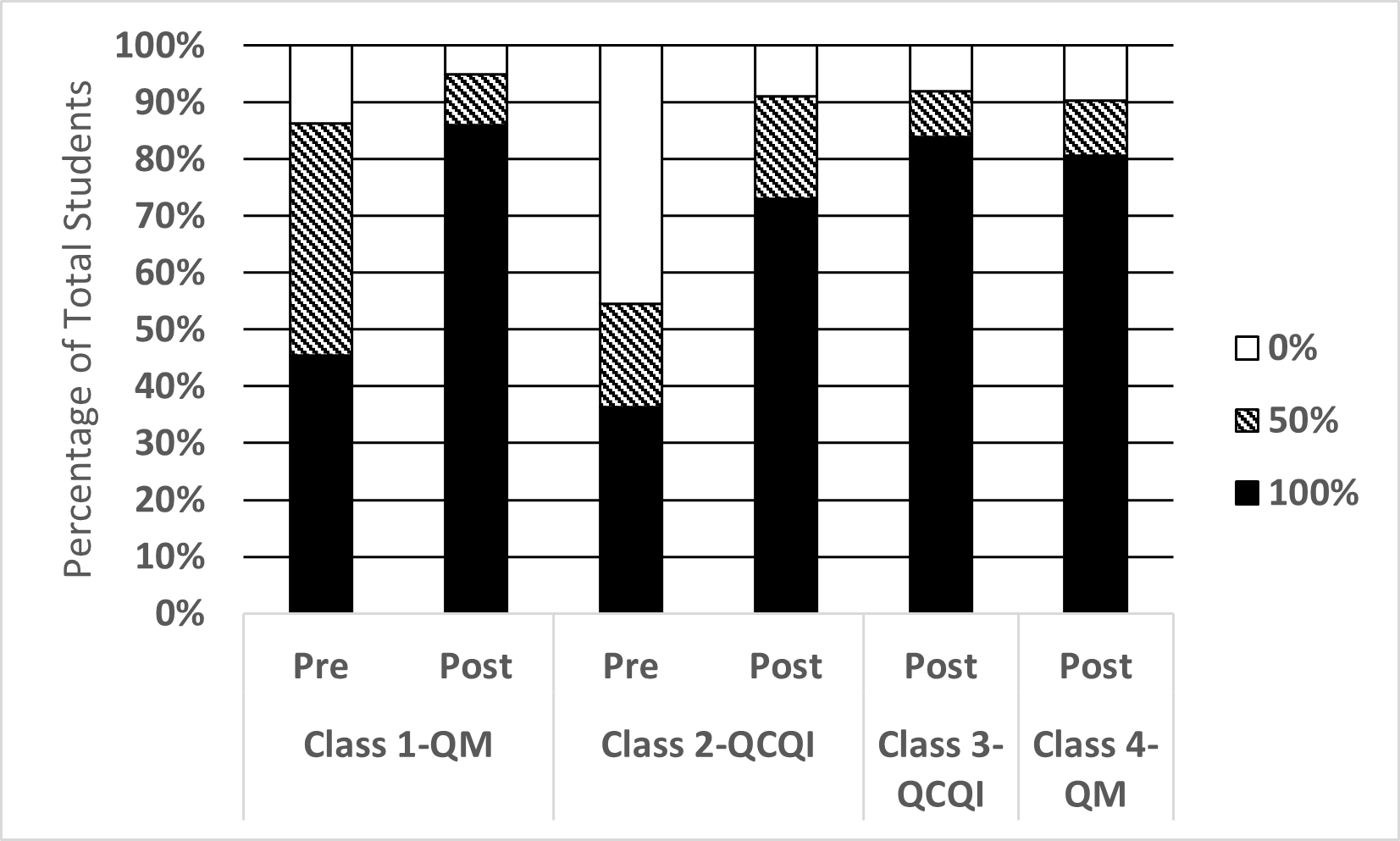}
        \caption[Q3]%
        {{\small Score breakdown for question 3.}}    
        \label{fig:Q3_Breakdown}
    \end{subfigure}
    \hfill
    \begin{subfigure}[b]{0.475\textwidth}   
        \centering 
        \includegraphics[width=\textwidth]{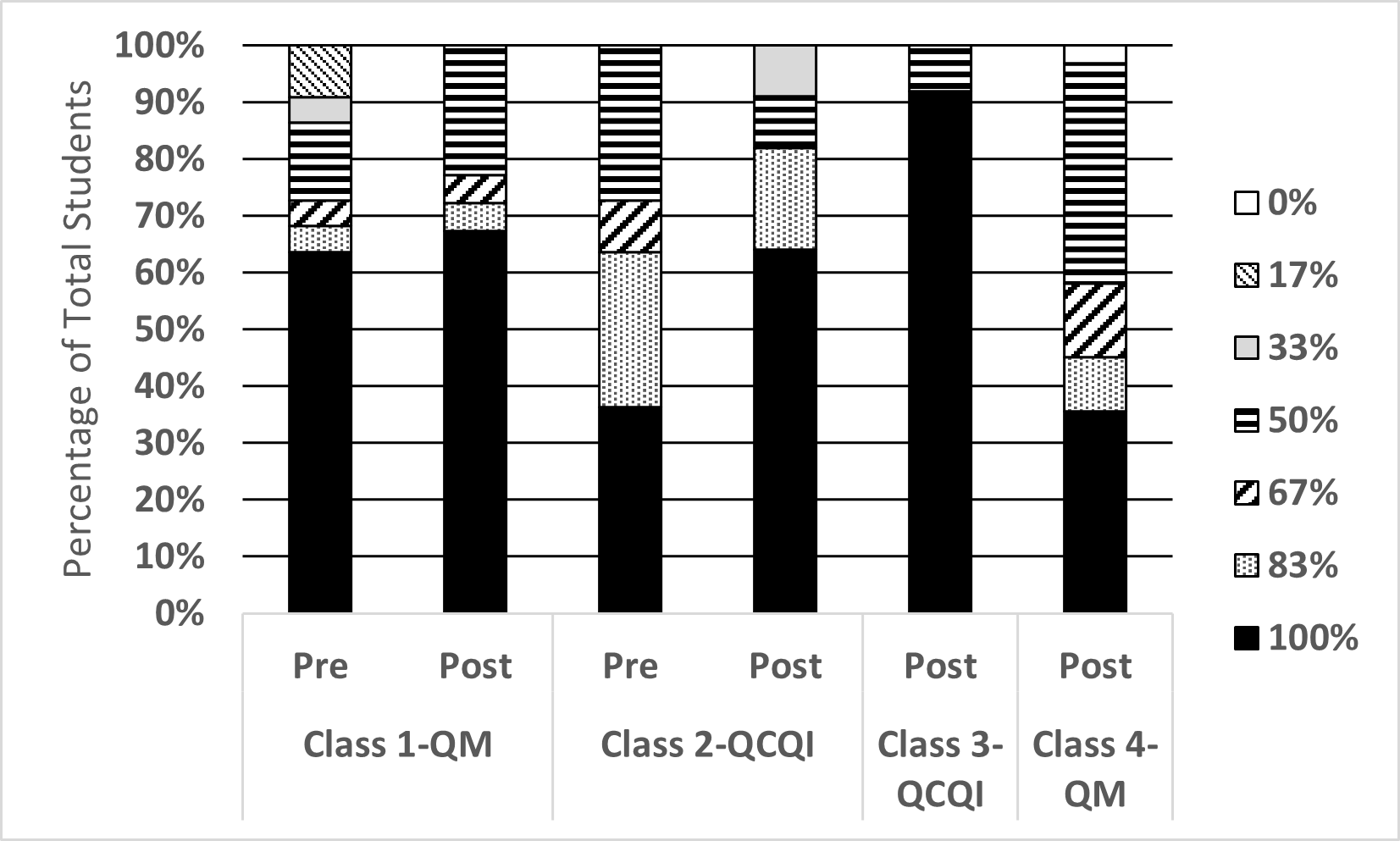}
        \caption[Q4]%
        {{\small Score breakdown for question 4.}}    
        \label{fig:Q4_Breakdown}
    \end{subfigure}
    \caption[]
    {\small The score breakdowns for each question, showing what percentage of students got a given score for each question. Percentage scores are based on the rubric provided in Section \ref{DataAnalysis}.} 
    \label{fig:all_Grade_Breakdowns}
\end{figure*}

Results from Table \ref{table:QCQI_Comparison} and Figure \ref{fig:all_Grade_Breakdowns} also help us compare, e.g., how students in the QCQI course did after engaging with the topics for the first time regardless of whether they had lecture-based instruction or engaged with the tutorial on QKD.
Table \ref{table:QCQI_Comparison} shows that the QCQI courses performed reasonably well on Q1 and Q2 after engaging with the QKD concepts (involving entanglement) for the first time (whether via lecture or the tutorial). This is encouraging as it emphasizes that even students in an interdisciplinary course are able to understand how Alice and Bob's strategy works to create their shared key regardless of initial instructional method. For Q3, students in Class 2-QCQI performed poorly on the pre-test after traditional lecture on QKD (45\%) but after engaging with the tutorial, Class 2-QCQI's average score is 82\% which is comparable to Class 3-QCQI's average score of 88\%. Thus, Table \ref{table:QCQI_Comparison} shows that both Class 2-QCQI and Class 3-QCQI did equally well on Q3 after both classes engaged with the tutorial (regardless of whether there was a lecture on QKD). For Q4, the average score of Class 2-QCQI was 79\% after a traditional lecture on QKD and it improved to 86\% after engaging with the tutorial, while students in Class 3-QCQI who only engaged with the tutorial scored 96\%. However, this difference in performance for Class 2-QCQI and Class 3-QCQI after the tutorial may be due to the difference in emphasis on Bell states in each class since the instructors were different (apart from individual differences among students since class sizes are small). 

\subsection{Response from a Student who did not engage with the QuILT}

One of the students from Class 4-QM took the post-test without engaging with the QuILT and submitting it as homework. While we did not include their data in in-class results presented here in Tables, we found that their response gave another look at the efficacy of the QuILT.

This student was able to correctly answer Q1 and Q2 on the post-test since the protocol for QKD was provided as part of the assessment. However, in response to Q3, this student stated ``Eve's measurements would change the data received by Alice and Bob. In the given example, if Eve measures a 1 along the [Z-axis] and sends it back to Bob, his `uncertain' results would become quite consistent if she sends a particle in the same state every time." We interpreted this as the student potentially stating that, somehow, Bob will continually receive the same state from Eve and that would cause her to be detected. Finally in response to Q4, this student selected choice E (which is the state $\frac{1}{\sqrt{2}} (\ket{\uparrow_{A}} \ket{\downarrow_{B}} - \ket{\downarrow_{A}}\ket{\downarrow_{B}})$) and simply wrote ``= $\ket{\Psi_{AB}}$". 

We found these responses interesting as, although this student was able to answer Q1 and Q2 correctly based upon the information about the protocol between Alice and Bob provided with the test, they did not have an understanding of the underlying concepts that guide QKD. This student correctly answering Q1 and Q2 implies that our description of the protocol provided to students at the beginning of the pre-/post-test is enough to give students an understanding of the protocol. Their response to Q3 did not include anything relating to which basis Alice and Bob were in (Alice and Bob must be in the same basis but Eve should be in a different basis for Eve's presence to be detected since that is the case in which they include the information in their key) and appears to imply that Eve's measurement will cause some amount of consistency in Bob's measurements. Their response to Q4 implies that they do not understand the importance of entanglement to the QKD protocol. We found this to be encouraging as it shows that the QuILT provides students with a reasonably good understanding of QKD involving entanglement. 

\subsection{Retention and Consolidation} \label{Retention}

Approximately one month after the post-test was administered to the second session of Class 4-QM, they were given a final exam which included the post-test again to investigate retention and consolidation of learning one month later. At this point, students had received their prior post-test back with feedback on their responses (note that they had not explicitly been informed that the 
QKD post-test would be given again as part of the final exam but they were generally informed that the final exam was a cumulative exam and could include anything they had learned from the entire semester). Results are shown in Table \ref{table:initial_vs_retention} below.

\begin{table}[ht]
    \centering
    \begin{tabular}{|c|c|c|}
    \hline
       \multirow{2}{*}{\textbf{Question Number}} &
      \multicolumn{1}{c|}{\textbf{Class 4-QM initial post-test}} &
      \multicolumn{1}{c|}{\textbf{Class 4-QM retention post-test}}\\
     & Post & Post \\
    \hline
    1 & 97\% & 100\%\\
    2 & 96\% & 96\%\\
    3 & 85\% & 83\%\\
    4 & 72\% & 90\%\\   
    \hline
    \end{tabular}
    \caption{Comparison of the initial post-test scores from Class 4-QM to the post-test which was included in the second session of Class 4-QM's final exam (retention post-test).}
    \label{table:initial_vs_retention}
\end{table}

We note that since this retention post-test was only given to one of the two sessions included in Class 4-QM, there are only 22 students whose data are included (versus the 31 students included in the data for the initial post-test which includes both sessions from Class 4-QM). For Q1, Q2, and Q3, we see marginal changes, implying that students had good retention of the concepts relating to both how the protocol works when only Alice and Bob are present (Q1 and Q2) and concepts relating to how the process guarantees the detection of an eavesdropper's presence (Q3). This is encouraging as it shows that students were able to retain what they had learned from the QuILT after a month.
Interestingly, for Q4 student performance improved. This may be due to the fact that students received feedback on their initial post-test and learned from it. This is encouraging as it implies that not only did students do reasonably well on this transfer question after they engaged with the QuILT (72\% average), but many learned from their mistakes and improved on the retention post-test (90\%). We note that this question involves correctly identifying the correct multiple choice options (which was given as feedback to each student on their initial post-test) as well as correctly describing the reason behind the multiple choice questions being correct (which was not given explicitly in the feedback for each student's initial post-test).

\subsection{Limitations of Study and Outlook} \label{Limitations}

It is important to note that the QKD method presented here, which uses two SGAs, is unlikely to be utilized in QKD real-world scenarios. Instead, we are likely to see the BB84 protocol or other protocols noted in Section \ref{Intro} \cite{bennett1984,ekert1991quantum,kwiatqkd,RevModPhysqkd,qkdpanchina2017}, which usually utilize polarization states of single photons. However, we find that the protocol presented here, despite it's limited real world use, can be a helpful step to understanding these more common QKD protocols. This was seen in interviews when the interviewees were asked to engage with a different QuILT, which utilized a different QKD method \cite{devore2020qkd}. While the interviewees completed this second QuILT, we found that they tended to draw parallels between both methods in order to assist their understanding. 

We also note that in the real-world use of QKD, considerations of noise which occurs in these quantum systems, especially over large distances, must be considered. Though not discussed in the QuILT, it may be helpful framing for students who are interested in entering the QISE field to understand some of the challenges they will be facing, or alternatively it could be framed as difficulties which are relevant to many aspects of the second quantum revolution as a whole.

We also note that due to the nature of giving the QuILT as a homework, it may not be possible to ensure a level of collaborative engagement with peers with the interactivity and engagement of instructor for overall feedback after certain number of questions.

The QuILT and pre-/post-test was designed to be a tool which would not infringe greatly on valuable in-class lecture time. We note that the brevity of the pre-/post-test could limit the generalizability of these results. It is possible that certain alternative conceptions which were not addressed could be persistent after student engagement with the QuILT without peer discussion and overall class discussion by instructor after a certain number of questions.

Another potential limitation is that the pre-/post-test was not given to students who had not yet learned about QKD in some capacity. It may be an interesting future investigation to study how well students are able to perform on the pre-/post-test before learning about QKD to act as a further control group.

After utilizing this QuILT, relevant outlooks to next topics may be Bell measurements, other QKD methods that are more commonly used (such as BB84 or E91), or this QuILT may be a helpful step towards other relevant topics which utilize entanglement.

\vspace*{-0.2in}
\section{Summary and Implications for Teaching} \label{Discuss}
\vspace*{-0.1in}

We find that students are able to more accurately describe and explain the way that the QKD protocol involving entanglement allows two parties to generate a shared secure key after engaging with the tutorial vs. after lecture-based instruction. 
Also, students can learn QKD concepts in the context of an application by engaging with the tutorial as homework after learning about entanglement even if there was no lecture-based instruction on the QKD concepts before engaging with the tutorial. This suggests that the tutorial could be effective as a homework that is given to students after they are taught about entanglement without traditional lecture-based instruction on QKD.

Our findings also suggest that the tutorial is effective in helping students, both in a QM course and a foundations of quantum computing and quantum information course (for an interdisciplinary student body), apply their knowledge of entanglement in a physical QKD context relevant for quantum cryptography. With this in mind, the tutorial can be effectively used as follows:

\begin{itemize}
    \item The tutorial could be helpful as a homework assignment to help students learn about QKD after learning about entanglement in class, as seen from performances of 
    Class 3-QCQI and Class 4-QM. We note that the tutorial's efficacy may be due to the fact that QKD using entanglement is an excellent way to physically connect the concept of entanglement to a real world application in cryptography. We believe this would allow instructors to circumvent having to utilize in-class time to teach about QKD via lecture-based instruction, but still allow them to include the concept as homework with a similar depth as they would have achieved if they had done so in class.
    
    \item The tutorial could be used in part or in full in class. We believe this would be best done after the instructor teaches entanglement including Bell states, as the QKD protocol is reliant upon entanglement. While students can work in small groups on the tutorial, the instructor can move around to ensure that students are engaging productively with each other and the tutorial. Although we suggest utilizing the whole tutorial to ensure full understanding, if there is not enough time, it may be helpful to only utilize parts of the tutorial based upon instructor preference for in-class discussion on these issues. 

    \item After students learn about entanglement, questions in the tutorial could be framed as think-pair-share or clicker questions. This option allows for instructors to obtain feedback on the difficulties that their classes are experiencing to fine-tune instruction further to their students' needs.
\end{itemize}

The tutorial is designed to allow instructors flexibility in using their preferred methods for instruction, or no instruction if they give it as homework only. Thus, instructors can use the tutorial in part or in full, based on their preference, as a resource either in or out of the classroom to assist with student understanding of these QKD concepts relevant for QISE.

\section*{Funding}
This research is supported by the US National Science Foundation Award PHY-2309260.

\section*{Informed Consent Statement}

The research was approved as exempt by Institutional Review Board (IRB) and informed consent was obtained from interviewed students.

\section*{Availability of data and materials}
The datasets used and analyzed during the current study are not available based on IRB protocol and confidentiality agreement.

\section*{Authors' contributions}

L.D. and C.S. contributed to analysis and interpretation of data, as well as writing and revision of the manuscript. C.S. contributed to the conception and design of research. L.D. and C.S. contributed to acquisition and interpretation of data, and revision of the manuscript. All authors read and approved the final manuscript.

\section*{Ethical Statement}

This research was carried out in accordance with the principles outlined in the university’s Institutional Review Board ethical policy.

\vspace*{-0.2in}
\section{Appendix: The Pre and Post-Test}
\vspace*{-0.1in}

1. Assume Alice and Bob conduct a large number of measurements using the protocol discussed. Complete the following table by recording in the third column whether or not Bob knows with certainty what Alice measures from the knowledge of Alice’s SGA orientation and knowledge of his own measurement:

\begin{figure}[h!]
    \centering
    \includegraphics[scale = 1]{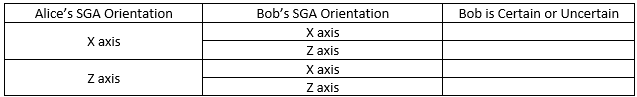}
    \label{fig:prepostQ1}
\end{figure}

2. Complete the following table with the measurements made by Bob in each case (write 1, 0, or “–'' where you are not certain because it could be either 1 or 0).

\begin{figure}[h!]
    \centering
    \includegraphics[scale = 1]{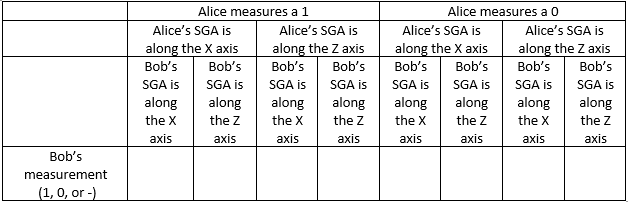}
    \label{fig:prepostQ2}
\end{figure}

3. Now let us assume that there is a third party, Eve, who intercepts every particle in the entangled state being sent to Bob with her own SGA which she randomly orients along the X or Z direction (like the strategy used by Alice and Bob). Then, after intercepting a particle, Eve immediately generates a replacement particle (unlike the entangled state) to send to Bob in its place. Let us assume that Eve can generate the replacement particle instantly and that she matches the particle that she intercepted (e.g., if Eve measures a 1 along the Z axis, she produces and sends a particle to Bob which is also in the $\ket{\uparrow_{B}}_{Z}$ state). Once the random shared encryption and decryption key has been generated bit by bit using the above protocol, a small subset of bits, e.g., every 10th bit, is compared by Alice and Bob over a public channel to make sure they are consistent (they discard these bits that they compare so that these bits are not part of the random shared key generated). In not more than three sentences, explain why Eve’s presence would be detected by Alice and Bob (cite specific cases in which her presence would be detected).

\medskip

4. In a quantum key distribution protocol which uses entangled pairs, which of the following quantum states will be equally suitable for quantum key distribution? Explain why.

    (a)  $\frac{1}{\sqrt{2}} (\ket{\uparrow_{A}} \ket{\uparrow_{B}} + \ket{\downarrow_{A}}\ket{\downarrow_{B}})$
    
    (b) $\frac{1}{\sqrt{2}} (\ket{\uparrow_{A}} \ket{\downarrow_{B}} + \ket{\downarrow_{A}}\ket{\uparrow_{B}})$
    
    (c) $\frac{1}{\sqrt{2}} (\ket{\uparrow_{A}} \ket{\uparrow_{B}} + \ket{\uparrow_{A}}\ket{\downarrow_{B}})$
    
    (d) $\frac{1}{\sqrt{2}} (\ket{\uparrow_{A}} \ket{\uparrow_{B}} - \ket{\downarrow_{A}}\ket{\downarrow_{B}})$
    
    (e) $\frac{1}{\sqrt{2}} (\ket{\uparrow_{A}} \ket{\downarrow_{B}} - \ket{\downarrow_{A}}\ket{\downarrow_{B}})$

\medskip

\textbf{Answer Key:}

1. Certain, Uncertain, Uncertain, Certain

2. 0, -, -, 0, 1, -, -, 1

3. Student responses must acknowledge that Eve must be in a different basis than Alice and Bob as well as acknowledging that, even in the case in which Eve is in a different basis than Alice and Bob, the detection of Eve is still probabilistic in nature (e.g. students cannot say that Alice and Bob would identify Eve immediately once she is in a different basis than they are, students must acknowledge that Alice and Bob need to check multiple bits to check for discrepancies).

4. Choices A, B, and D should be selected and these states are all suitable because, if Alice and Bob are in the same basis, when Alice makes a measurement in that basis it guarantees only one outcome for Bob's measurement (Alternatively, answers along the lines of ``These states [choices A, B, and D] are Bell States/maximally entangled" were considered acceptable as they showed an understanding that the QKD method requires maximally entangled states).

\bibliography{refs}

@incollection{zuza2023teachinglearning,
    author = {Guisasola, Jenaro and Zuza, Kristina and Sarriugarte, Paulo and Ametller, Jaume},
    editor = {Taşar, Mehmet Fatih and Heron, Paula R. L.},
    isbn = {978-0-7354-2548-4},
    title = {Research-Based Teaching-Learning Sequences in Physics Education: A Rising Line of Research},
    booktitle = {The International Handbook of Physics Education Research: Special Topics},
    publisher = {AIP Publishing LLC},
    doi = {10.1063/9780735425514_026},
    url = {https://doi.org/10.1063/9780735425514\_026},
    eprint = {\url{https://pubs.aip.org/book/chapter-pdf/18252854/9780735425514_026.pdf}},
}

@book{wong2022introduction,
  title={Introduction to Classical and Quantum Computing},
  author={Wong, Thomas G},
  year={2022},
  publisher={Rooted Grove Omaha, NE, USA}
}

@book{townsend2000modern,
  title={A Modern Approach to Quantum Mechanics},
  author={Townsend, John S},
  year={2000},
  publisher={University Science Books}
}

@book{griffithsQM,
  title={Introduction to Quantum Mechanics},
  author={Griffiths, David J and Schroeter, Darrell F},
  year={2018},
  publisher={Cambridge university press}
}

@misc{NSAqkdsecurity,
    title = {Quantum Key Distribution (QKD) and Quantum Cryptography (QC)},
    author = {National Security Agency/Central Security Service},
    howpublished = {\url{https://www.nsa.gov/Cybersecurity/Quantum-Key-Distribution-QKD-and-Quantum-Cryptography-QC/}}
}

@article{B92protocol,
  title = {Quantum cryptography using any two nonorthogonal states},
  author = {Bennett, Charles H.},
  journal = {Phys. Rev. Lett.},
  volume = {68},
  issue = {21},
  pages = {3121--3124},
  numpages = {0},
  year = {1992},
  month = {May},
  publisher = {American Physical Society},
  doi = {10.1103/PhysRevLett.68.3121},
  url = {https://link.aps.org/doi/10.1103/PhysRevLett.68.3121}
}

@article{bennett1992quantum,
  title={Quantum cryptography without {B}ell’s theorem},
  author={Bennett, Charles H and Brassard, Gilles and Mermin, N David},
  journal={Physical Review Letters},
  volume={68},
  number={5},
  pages={557},
  year={1992},
  publisher={APS}
}

@Article{justicemathphysics,
AUTHOR = {Justice, Paul D. and Marshman, Emily and Singh, Chandralekha},
TITLE = {A Framework for Understanding the Impact of Integrating Conceptual and Quantitative Reasoning in a Quantum Optics Tutorial on Students’ Conceptual Understanding},
JOURNAL = {Education Sciences},
VOLUME = {15},
YEAR = {2025},
NUMBER = {10},
ARTICLE-NUMBER = {1314},
URL = {https://www.mdpi.com/2227-7102/15/10/1314},
ISSN = {2227-7102},
DOI = {10.3390/educsci15101314}
}

@article{nocloning,
  title={A single quantum cannot be cloned},
  author={Wootters, William K and Zurek, Wojciech H},
  journal={Nature},
  volume={299},
  number={5886},
  pages={802--803},
  year={1982},
  publisher={Nature Publishing Group UK London}
}

@book{woottersqkd,
  title={Protecting Information: From Classical Error Correction to Quantum Cryptography},
  author={Loepp, Susan and Wootters, William K},
  year={2006},
  publisher={Cambridge University Press}
}

@article{qkdpanchina2017,
   author = {Liao, Sheng-Kai and Cai, Wen-Qi and others},
   title = {Satellite-to-ground quantum key distribution},
   journal = {Nature},
   volume = {549},
   number = {7670},
   pages = {43-47},
   year = {2017},
   type = {Journal Article}
}

@article{qtmerzel,
   author = {Weissman, E.Y. and Merzel, A. and Katz, N. and Galili, I. },
   title = {Keep it secret, keep it safe: Teaching quantum key distribution in high school},
   journal = {EPJ Quantum Technol.},
   volume = {11},
   pages = {64},
   url = {https://doi.org/10.1140/epjqt/s40507-024-00276-4},
   year = {2024},
   type = {Journal Article}
}

@article{kellyqkd,
    author = {Schneble, Dominik and Wei, Tzu-Chieh and Kelly, Angela M.},
    title = {Quantum information science and technology high school outreach: Conceptual progression for introducing principles and programming skills},
    journal = {American Journal of Physics},
    volume = {93},
    number = {1},
    pages = {88-97},
    year = {2025},
    month = {01},
    issn = {0002-9505},
    doi = {10.1119/5.0211535},
    url = {https://doi.org/10.1119/5.0211535},
    
}

@article{netoqkdoutreach,
    author = {Neto Mendes, Pedro and André, Paulo and Zambrini Cruzeiro, Emmanuel},
    title = {Simple portable quantum key distribution for science outreach},
    journal = {American Journal of Physics},
    volume = {93},
    number = {1},
    pages = {69-77},
    year = {2025},
    month = {01},
    issn = {0002-9505},
    doi = {10.1119/5.0204077},
    url = {https://doi.org/10.1119/5.0204077},

}

@article{kwiatqkd,
  title = {Entangled State Quantum Cryptography: Eavesdropping on the {E}kert Protocol},
  author = {Naik, D. S. and Peterson, C. G. and White, A. G. and Berglund, A. J. and Kwiat, P. G.},
  journal = {Phys. Rev. Lett.},
  volume = {84},
  issue = {20},
  pages = {4733--4736},
  numpages = {0},
  year = {2000},
  month = {May},
  publisher = {American Physical Society},
  doi = {10.1103/PhysRevLett.84.4733},
  url = {https://link.aps.org/doi/10.1103/PhysRevLett.84.4733}
}

@article{RevModPhysqkd,
  title = {The security of practical quantum key distribution},
  author = {Scarani, Valerio and Bechmann-Pasquinucci, Helle and others},
  journal = {Rev. Mod. Phys.},
  volume = {81},
  issue = {3},
  pages = {1301--1350},
  numpages = {0},
  year = {2009},
  month = {Sep},
  publisher = {American Physical Society},
  doi = {10.1103/RevModPhys.81.1301},
  url = {https://link.aps.org/doi/10.1103/RevModPhys.81.1301}
}

@article{Kohnleqkd,
doi = {10.1088/1361-6404/aa62c8},
url = {https://dx.doi.org/10.1088/1361-6404/aa62c8},
year = {2017},
month = {mar},
publisher = {IOP Publishing},
volume = {38},
number = {3},
pages = {035403},
author = {Kohnle, Antje and Rizzoli, Aluna},
title = {Interactive simulations for quantum key distribution},
journal = {European Journal of Physics}
}

@article{galvezqkd,
    author = {Bista, Aayam and Sharma, Baibhav and Galvez, Enrique J.},
    title = {A demonstration of quantum key distribution with entangled photons for the undergraduate laboratory},
    journal = {American Journal of Physics},
    volume = {89},
    number = {1},
    pages = {111-120},
    year = {2021},
    month = {01},
    issn = {0002-9505},
    doi = {10.1119/10.0002169},
    url = {https://doi.org/10.1119/10.0002169},
    
}

@misc{quantique,
author = {IDQuantique},
note ={https://www.idquantique.com/},
url = {https://www.idquantique.com/},
doi = {https://www.idquantique.com/}
}

@book{taskkirwan,
author = {Kirwan, B. and Ainsworth, L. K.},
title = {A Guide to Task Analysis},
publisher = {Taylor \& Francis},
address = {London; Washington, DC},
ISBN = {0748400575 9780748400577 0748400583 9780748400584},
year = {1992},
type = {Book}
}

@book{vygotsky,
author = {Vygotsky, Lev Semenovich and Cole, Michael},
title = {Mind in Society: The Development of Higher Psychological Processes},
publisher = {Harvard University Press},
ISBN = {0674576292},
year = {1978},
type = {Book}
}

@article{schwartzbransford,
author = {Schwartz, Daniel L and Bransford, John D and Sears, David},
title = {Efficiency and innovation in transfer},
journal = {Transfer of learning from a modern multidisciplinary perspective},
volume = {3},
number = {1},
pages = {1-51},
year = {2005},
type = {Journal Article}
}

@article{schwartz1998time,
author = {Schwartz, Daniel L. and Bransford, John D.},
title = {A time for telling},
journal = {Cognition and instruction},
volume = {16},
number = {4},
pages = {475--5223},
year = {1998},
type = {Journal Article}
}

@article{marshman2016ejpphoton,
author = {Marshman, Emily and Singh, Chandralekha},
title = {Interactive tutorial to improve student understanding of single photon experiments involving a {M}ach–{Z}ehnder interferometer},
journal = {European Journal of Physics},
volume = {37},
number = {2},
pages = {024001},
ISSN = {0143-0807},
year = {2016},
type = {Journal Article}
}

@article{maries2020mzidouble,
author = {Maries, Alexandru and Sayer, Ryan and Singh, Chandralekha},
title = {Can students apply the concept of “which-path” information learned in the context of {M}ach–{Z}ehnder interferometer to the double-slit experiment?},
journal = {American Journal of Physics},
volume = {88},
number = {7},
pages = {542-550},
ISSN = {0002-9505},
DOI = {10.1119/10.0001357},
url = {http://dx.doi.org/10.1119/10.0001357},
year = {2020},
type = {Journal Article}
}

@article{Benlarmorajp2025,
    author = {Brown, Ben and Zhu, Guangtian and Singh, Chandralekha},
    title = {Investigating and improving student understanding of time dependence of expectation values in quantum mechanics using an interactive tutorial on {L}armor precession},
    journal = {American Journal of Physics},
    volume = {93},
    number = {1},
    pages = {52-57},
    year = {2025},
    month = {01},
    issn = {0002-9505},
    doi = {10.1119/5.0186030},
    url = {https://doi.org/10.1119/5.0186030}
}

@article{asfaw2022ieee,
   author = {Asfaw, A. and Blais, A. and others},
   title = {Building a Quantum Engineering Undergraduate Program},
   journal = {IEEE Transactions on Education},
   volume = {65},
   number = {2},
   pages = {220-242},
   ISSN = {1557-9638},
   DOI = {10.1109/TE.2022.3144943},
   year = {2022},
   type = {Journal Article}
}

@inproceedings{bennett1984,
  title={Quantum cryptography: Public key distribution and coin tossing},
  author={Bennett, Charles H and Brassard, Gilles},
  booktitle={Proceedings of the International Conference on Computers, Systems and Signal Processing},
  pages={175--179},
  year={1984}
}

@article{devore2020qkd,
   author = {DeVore, Seth and Singh, Chandralekha},
   title = {Interactive learning tutorial on quantum key distribution},
   journal = {Physical Review Physics Education Research},
   volume = {16},
   number = {1},
   pages = {010126},
   DOI = {10.1103/PhysRevPhysEducRes.16.010126},
   url = {https://link.aps.org/doi/10.1103/PhysRevPhysEducRes.16.010126},
   year = {2020},
   type = {Journal Article}
}

@article{fox2020cu,
   author = {Fox, Michael F. J. and Zwickl, Benjamin M. and Lewandowski, H. J.},
   title = {Preparing for the quantum revolution: What is the role of higher education?},
   journal = {Physical Review Physics Education Research},
   volume = {16},
   number = {2},
   pages = {020131},
   DOI = {10.1103/PhysRevPhysEducRes.16.020131},
   url = {https://link.aps.org/doi/10.1103/PhysRevPhysEducRes.16.020131},
   year = {2020},
   type = {Journal Article}
}

@article{muller2023prperworkforce,
   author = {Greinert, Franziska and Müller, Rainer and Bitzenbauer, Philipp and Ubben, Malte S. and Weber, Kim-Alessandro},
   title = {Future quantum workforce: Competences, requirements, and forecasts},
   journal = {Physical Review Physics Education Research},
   volume = {19},
   number = {1},
   pages = {010137},
   DOI = {10.1103/PhysRevPhysEducRes.19.010137},
   url = {https://link.aps.org/doi/10.1103/PhysRevPhysEducRes.19.010137},
   year = {2023},
   type = {Journal Article}
}

@article{kohnle2013,
   author = {Kohnle, Antje and Bozhinova, Inna and Browne, Dan and Everitt, Mark and Fomins, Aleksejs and Kok, Pieter and Kulaitis, Gytis and Prokopas, Martynas and Raine, Derek and Swinbank, Elizabeth},
   title = {A new introductory quantum mechanics curriculum},
   journal = {European Journal of Physics},
   volume = {35},
   number = {1},
   pages = {015001},
   ISSN = {0143-0807},
   year = {2013},
   type = {Journal Article}
}

@article{marshman2015,
   author = {Marshman, Emily and Singh, Chandralekha},
   title = {Framework for understanding the patterns of student difficulties in quantum mechanics},
   journal = {Physical Review Special Topics-Physics Education Research},
   volume = {11},
   number = {2},
   pages = {020119},
   year = {2015},
   type = {Journal Article}
}

@article{marshman2017opejp,
   author = {Marshman, Emily and Singh, Chandralekha},
   title = {Investigating and improving student understanding of quantum mechanical observables and their corresponding operators in {Dirac} notation},
   journal = {European Journal of Physics},
   volume = {39},
   number = {1},
   pages = {015707},
   year = {2017},
   type = {Journal Article}
}

@article{marshman2017expect,
   author = {Marshman, Emily and Singh, Chandralekha},
   title = {Investigating and improving student understanding of the expectation values of observables in quantum mechanics},
   journal = {European Journal of Physics},
   volume = {38},
   number = {4},
   pages = {045701},
   year = {2017},
   type = {Journal Article}
}

@article{marshman2017prob,
   author = {Marshman, Emily and Singh, Chandralekha},
   title = {Investigating and improving student understanding of the probability distributions for measuring physical observables in quantum mechanics},
   journal = {European Journal of Physics},
   volume = {38},
   number = {2},
   pages = {025705},
   year = {2017},
   type = {Journal Article}
}

@article{marshman2019qmfps,
   author = {Marshman, Emily and Singh, Chandralekha},
   title = {Validation and administration of a conceptual survey on the formalism and postulates of quantum mechanics},
   journal = {Physical Review Physics Education Research},
   volume = {15},
   number = {2},
   pages = {020128},
   DOI = {https://link.aps.org/doi/10.1103/PhysRevPhysEducRes.15.020128},
   url = {https://link.aps.org/doi/10.1103/PhysRevPhysEducRes.15.020128},
   year = {2019},
   type = {Journal Article}
}

@article{meyer2022cu,
   author = {Meyer, Josephine C. and Passante, Gina and Pollock, Steven J. and Wilcox, Bethany R.},
   title = {Today's interdisciplinary quantum information classroom: Themes from a survey of quantum information science instructors},
   journal = {Physical Review Physics Education Research},
   volume = {18},
   number = {1},
   pages = {010150},
   DOI = {10.1103/PhysRevPhysEducRes.18.010150},
   url = {https://link.aps.org/doi/10.1103/PhysRevPhysEducRes.18.010150},
   year = {2022},
   type = {Journal Article}
}

@article{cedricrepresentation2012,
doi = {10.1088/0143-0807/33/3/657},
url = {https://doi.org/10.1088/0143-0807/33/3/657},
year = {2012},
month = {mar},
publisher = {IOP Publishing},
volume = {33},
number = {3},
pages = {657},
author = {Fredlund, Tobias and Airey, John and Linder, Cedric},
title = {Exploring the role of physics representations: an illustrative example from students sharing knowledge about refraction},
journal = {European Journal of Physics}
}

@article{representationlin2017,
  title = {Challenges in designing appropriate scaffolding to improve students' representational consistency: The case of a {G}auss's law problem},
  author = {Maries, Alexandru and Lin, Shih-Yin and Singh, Chandralekha},
  journal = {Phys. Rev. Phys. Educ. Res.},
  volume = {13},
  issue = {2},
  pages = {020103},
  numpages = {17},
  year = {2017},
  month = {Aug},
  publisher = {American Physical Society},
  doi = {10.1103/PhysRevPhysEducRes.13.020103},
  url = {https://link.aps.org/doi/10.1103/PhysRevPhysEducRes.13.020103}
}

@article{representation2024,
  title = {Using multiple representations to improve student understanding of quantum states},
  author = {Marshman, Emily and Maries, Alexandru and Singh, Chandralekha},
  journal = {Phys. Rev. Phys. Educ. Res.},
  volume = {20},
  issue = {2},
  pages = {020152},
  numpages = {16},
  year = {2024},
  month = {Dec},
  publisher = {American Physical Society},
  doi = {10.1103/PhysRevPhysEducRes.20.020152},
  url = {https://link.aps.org/doi/10.1103/PhysRevPhysEducRes.20.020152}
}

@article{michelini2022,
   author = {Michelini, Marisa and Stefanel, Alberto and Tóth, Kristóf},
   title = {Implementing {Dirac} Approach to Quantum Mechanics in a {Hungarian} Secondary School},
   journal = {Education Sciences},
   volume = {12},
   number = {9},
   pages = {606},
   ISSN = {2227-7102},
   url = {https://doi.org/10.3390/educsci12090606},
   year = {2022},
   type = {Journal Article}
}

@article{raymer2019,
   author = {Raymer, Michael G. and Monroe, Christopher},
   title = {The {US} national quantum initiative},
   journal = {Quantum Science and Technology},
   volume = {4},
   number = {2},
   pages = {020504},
   year = {2019},
   type = {Journal Article}
}

@article{singh2007comp,
   author = {Singh, Chandralekha},
   title = {Helping Students Learn Quantum Mechanics for Quantum Computing},
   journal = {AIP Conf. Proc.},
   volume = {883},
   number = {1},
   pages = {42-45},
   year = {2007},
doi={https://doi.org/10.1063/1.2508687},
url={https://doi.org/10.1063/1.2508687},
   type = {Journal Article}
}

@article{singhasfaw2021pt,
   author = {Singh, Chandralekha and Asfaw, Abraham and Levy, Jeremy},
   title = {Preparing students to be leaders of the quantum information revolution},
   journal = {Physics Today},
year = {2021},
DOI = {https://doi.org/10.1063/PT.6.5.20210927a},
   url = {https://physicstoday.scitation.org/do/10.1063/PT.6.5.20210927a/full/}
}

@article{singh2022tpt,
   author = {Singh, Chandralekha and Levy, Akash and Levy, Jeremy},
   title = {Preparing Precollege Students for the Second Quantum Revolution with Core Concepts in Quantum Information Science},
   journal = {The Physics Teacher},
   volume = {60},
   number = {8},
   pages = {639-641},
   DOI = {10.1119/5.0027661},
   url = {https://doi.org/10.1119/5.0027661},
   year = {2022},
   type = {Journal Article}
}

@article{singh2015review,
   author = {Singh, Chandralekha and Marshman, Emily},
   title = {Review of student difficulties in upper-level quantum mechanics},
   journal = {Physical Review Special Topics-Physics Education Research},
   volume = {11},
   number = {2},
   pages = {020117},
   year = {2015},
   type = {Journal Article}
}

@article{zhu2012measure2,
   author = {Zhu, Guangtian and Singh, Chandralekha},
   title = {Improving students' understanding of quantum measurement. {II.} {D}evelopment of research-based learning tools},
   journal = {Physical Review Special Topics-Physics Education Research},
   volume = {8},
   number = {1},
   pages = {010118},
   year = {2012},
   type = {Journal Article}
}

@book{cohen,
author = {Cohen, Jacob},
title = {Statistical Power Analysis for the Behavioral Sciences},
publisher = {L. Erlbaum Associates},
address = {Hillsdale, N.J.},
year = {1988},
type = {Book}
}

@article{dorisquantum2025year,
  title = {Improving student understanding of quantum measurement in infinite-dimensional Hilbert space using a research-based multiple-choice question sequence},
  author = {Li, Yangqiuting and Singh, Chandralekha},
  journal = {Phys. Rev. Phys. Educ. Res.},
  volume = {21},
  issue = {1},
  pages = {010104},
  numpages = {27},
  year = {2025},
  month = {Jan},
  publisher = {American Physical Society},
  doi = {10.1103/PhysRevPhysEducRes.21.010104},
  url = {https://link.aps.org/doi/10.1103/PhysRevPhysEducRes.21.010104}
}

@article{Kashyapmisinformation,
doi = {10.1088/1361-6552/adbeb1},
url = {https://dx.doi.org/10.1088/1361-6552/adbeb1},
year = {2025},
month = {apr},
publisher = {IOP Publishing},
volume = {60},
number = {3},
pages = {035024},
author = {Kashyap, Jaya Shivangani and Singh, Chandralekha},
title = {Strategies educators can use to counter misinformation related to the quantum information revolution},
journal = {Physics Education}
}

@article{ekert1991quantum,
  title={Quantum cryptography based on {B}ell’s theorem},
  author={Ekert, Artur K},
  journal={Physical Review Letters},
  volume={67},
  number={6},
  pages={661},
  year={1991},
  publisher={APS}
}

@article{ghimire2025reflections,
  title={Reflections of quantum educators on strategies to diversify the second quantum revolution},
  author={Ghimire, Apekshya and Singh, Chandralekha},
  journal={The Physics Teacher},
  volume={63},
  number={1},
  pages={35--39},
  year={2025},
  publisher={AIP Publishing}
}

@article{ghimire2025epj,
  title={Investigating high school and pre-high school teachers’ perceptions and experiences introducing quantum concepts: a survey of {Q}uan{T}ime and other quantum-related activities},
  author={Ghimire, Apekshya and Kashyap, Jaya Shivangani and Edwards, Emily and Franklin, Diana and Singh, Chandralekha},
  journal={EPJ Quantum Technology},
  volume={12},
  number={1},
  pages={89},
  year={2025},
  publisher={Springer}
}

@article{fargol,
doi = {10.1088/1361-6404/ae0200},
url = {https://doi.org/10.1088/1361-6404/ae0200},
year = {2025},
month = {sep},
publisher = {IOP Publishing},
volume = {46},
number = {5},
pages = {055709},
author = {Seifollahi, Fargol and Singh, Chandralekha},
title = {Preparing students for the quantum information revolution: interdisciplinary teaching, curriculum development, and advising in quantum information science and engineering},
journal = {European Journal of Physics}
}

@article{liam2,
title={Building Bridges in Quantum Information Science Education: Expert Perspectives on Interdisciplinary Teaching and the Evolution of a Common Language},
  author={Doyle, L. and Seifollahi, F. and Singh, C.},
  journal={European Physical Journal Quantum Technology},
 volume = {13},
  issue = {2},
 year = {2026}
}

@article{liam,
  title = {Do we have a quantum computer? Expert perspectives on current state and future prospects},
  author = {Doyle, Liam and Seifollahi, Fargol and Singh, Chandralekha},
  journal = {Phys. Rev. Phys. Educ. Res.},
  volume = {22},
  issue = {1},
  pages = {010101},
  numpages = {26},
  year = {2026},
  month = {Jan},
  publisher = {American Physical Society},
  doi = {10.1103/md6x-pfrr},
  url = {https://link.aps.org/doi/10.1103/md6x-pfrr}
}

@article{hypeliam,
doi = {10.1088/1361-6404/ae3ae6},
url = {https://doi.org/10.1088/1361-6404/ae3ae6},
year = {2026},
month = {feb},
publisher = {IOP Publishing},
volume = {47},
number = {2},
pages = {025705},
author = {Doyle, Liam and Seifollahi, Fargol and Singh, Chandralekha},
title = {Navigating hype, interdisciplinary collaboration, and industry partnerships in quantum information science and technology: perspectives from leading quantum educators},
journal = {European Journal of Physics}
}

@Article{weissmanphysics,
AUTHOR = {Weissman, Efraim Yehuda and Merzel, Avraham and Katz, Nadav and Galili, Igal},
TITLE = {Phenomena and principles: Presenting quantum physics in a high school curriculum},
JOURNAL = {Physics},
VOLUME = {4},
YEAR = {2022},
NUMBER = {4},
PAGES = {1299--1317},
URL = {https://www.mdpi.com/2624-8174/4/4/83},
ISSN = {2624-8174},
}

@Article{bitzenbauer3,
AUTHOR = {Bitzenbauer, Philipp and Veith, Joaquin M. and Girnat, Boris and Meyn, Jan-Peter},
TITLE = {Assessing engineering students’ conceptual understanding of introductory quantum optics},
JOURNAL = {Physics},
VOLUME = {4},
YEAR = {2022},
NUMBER = {4},
PAGES = {1180--1201},
URL = {https://www.mdpi.com/2624-8174/4/4/77},
ISSN = {2624-8174},
}

@Article{bondani,
AUTHOR = {Bondani, Maria and Chiofalo, Maria Luisa and others},
TITLE = {Introducing quantum technologies at secondary school level: Challenges and potential impact of an online extracurricular course},
JOURNAL = {Physics},
VOLUME = {4},
YEAR = {2022},
NUMBER = {4},
PAGES = {1150--1167},
URL = {https://www.mdpi.com/2624-8174/4/4/75},
ISSN = {2624-8174},
}

@article{jeremytpt,
    author = {Levy, Jeremy and Singh, Chandralekha},
    title = {Hands-On Quantum: Teaching Core Quantum Concepts with {B}loch Cubes},
    journal = {The Physics Teacher},
    volume = {63},
    number = {2},
    pages = {85-89},
    year = {2025},
    month = {02},
}

@article{jeremyajp,
    author = {Levy, Jeremy and Singh, Chandralekha},
    title = {Teaching quantum formalism and postulates to first-year undergraduates},
    journal = {American Journal of Physics},
    volume = {93},
    number = {1},
    pages = {46-51},
    year = {2025},
    month = {01}
}

@article{divincenzo,
author = {DiVincenzo, David P.},
title = {The physical implementation of quantum computation},
journal = {Fortschritte der Physik},
year = {2000},
volume = {48},
number = {9-11},
pages = {771-783}
}

@article{kiko,
    author = {Bista, Aayam and Sharma, Baibhav and Galvez, Enrique J.},
    title = "{A demonstration of quantum key distribution with entangled photons for the undergraduate laboratory}",
    journal = {American Journal of Physics},
    volume = {89},
    number = {1},
    pages = {111-120},
    year = {2021},

    url = {https://doi.org/10.1119/10.0002169},
    
}

@article{flagship,
  title={Europe shows first cards in €1-billion quantum bet},
  author={Castelvecchi, Davide},
  journal={Nature},
  volume={563},
  number={7729},
  pages={14--15},
  year={2018}
}

@article{beckphoton,
    author = {Thorn, J. J. and Neel, M. S. and others},
    title = "{Observing the quantum behavior of light in an undergraduate laboratory}",
    journal = {American Journal of Physics},
    volume = {72},
    number = {9},
    pages = {1210-1219},
    year = {2004},

    url = {https://doi.org/10.1119/1.1737397},
   
}

@article{Kohnle_2017,
url = {https://dx.doi.org/10.1088/1361-6404/aa62c8},
year = {2017},
month = {mar},
publisher = {IOP Publishing},
volume = {38},
number = {3},
pages = {035403},
author = {Antje Kohnle and Aluna Rizzoli},
title = {Interactive simulations for quantum key distribution},
journal = {European Journal of Physics},

}

@article{hennig2024new,
  title={A new teaching-learning sequence to promote secondary school students’ learning of quantum physics using {D}irac notation},
  author={Hennig, Fabian and T{\'o}th, Krist{\'o}f and F{\"o}rster, Moritz and Bitzenbauer, Philipp},
  journal={Physics Education},
  volume={59},
  number={4},
  pages={045007},
  year={2024},
  publisher={IOP Publishing}
}

@article{qthellstern,
   author = {Hellstern, G. and Hettel, J. and  Just, B.},
   title = {Introducing quantum information and computation to a broader audience with {MOOC}s at {O}pen{HPI}},
   journal = {EPJ Quantum Technol.},
   volume = {11},
   pages = {59},
   year = {2024}
}

@article{qtsun,
  title={From computing to quantum mechanics: Accessible and hands-on quantum computing education for high school students},
  author={Sun, Qihong and Zhou, Shuangxiang and others},
  journal={EPJ Quantum Technology},
  volume={11},
  number={1},
  pages={58},
  year={2024},
  publisher={Springer Berlin Heidelberg}
}

@article{qtbrang,
  title={Spooky action at a distance? A two-phase study into learners’ views of quantum entanglement},
  author={Brang, Michael and Franke, Helena and others},
  journal={EPJ Quantum Technology},
  volume={11},
  number={1},
  pages={33},
  year={2024},
  publisher={Springer Berlin Heidelberg}
}

@article{qtmerzeletal,
  title={The core of secondary level quantum education: A multi-stakeholder perspective},
  author={Merzel, Avraham and Bitzenbauer, Philipp and others},
  journal={EPJ Quantum Technology},
  volume={11},
  number={1},
  pages={27},
  year={2024},
  publisher={Springer Berlin Heidelberg}
}

@article{qtmeyercu,
   author = {Meyer, Josephine C and Passante, Gina and Pollock, Steven J and Wilcox, Bethany R},
   title = {Introductory quantum information science coursework at {US} institutions: Content coverage},
   journal = {EPJ Quantum Technol.},
   volume = {11},
   pages = {16},
   url = {https://doi.org/10.1140/epjqt/s40507-024-00226-0},
   year = {2024},
   type = {Journal Article}
}

@article{qtgoorney,
  title={The Quantum Technology Open Master: Widening access to the quantum industry},
  author={Goorney, Simon and Sarantinou, Matoula and Sherson, Jacob},
  journal={EPJ Quantum Technology},
  volume={11},
  number={1},
  pages={7},
  year={2024},
  publisher={Springer Berlin Heidelberg}
}

@article{european,
url = {https://dx.doi.org/10.1088/1367-2630/aad1ea},
year = {2018},
month = {aug},
publisher = {IOP Publishing},
volume = {20},
number = {8},
pages = {080201},
author = {Antonio Acín and Immanuel Bloch and others},
title = {The quantum technologies roadmap: A {E}uropean community view},
journal = {New Journal of Physics}
}

@article{hucomputing,
  title = {Investigating and improving student understanding of the basics of quantum computing},
  author = {Hu, Peter and Li, Yangqiuting and Singh, Chandralekha},
  journal = {Phys. Rev. Phys. Educ. Res.},
  volume = {20},
  issue = {2},
  pages = {020108},
  numpages = {24},
  year = {2024},
  month = {Aug},
  publisher = {American Physical Society},
  url = {https://link.aps.org/doi/10.1103/PhysRevPhysEducRes.20.020108}
}

@article{Hubloch,
url = {https://dx.doi.org/10.1088/1361-6404/ad2393},
year = {2024},
month = {feb},
publisher = {IOP Publishing},
volume = {45},
number = {2},
pages = {025705},
author = {Peter Hu and Yangqiuting Li and Roger S K Mong and Chandralekha Singh},
title = {Student understanding of the {B}loch sphere},
journal = {European Journal of Physics},
}

@article{nvcenter2,
  title={Coherent control of {NV}- centers in diamond in a quantum teaching lab},
  author={Sewani, Vikas K and Vallabhapurapu, Hyma H and others},
  journal={American Journal of Physics},
  volume={88},
  number={12},
  pages={1156--1169},
  year={2020},
  publisher={AIP Publishing}
}

@article{lopez2020encrypt,
  title={Encrypt me! A game-based approach to {B}ell inequalities and quantum cryptography},
  author={L{\'o}pez-Incera, Andrea and Hartmann, Andreas and D{\"u}r, Wolfgang},
  journal={European Journal of Physics},
  volume={41},
  number={6},
  pages={065702},
  year={2020},
  publisher={IOP Publishing}
}

@article{rodriguez2020designing,
  title={Designing inquiry-based learning environments for quantum physics education in secondary schools},
  author={Rodriguez, Luiza Vilarta and van der Veen, Jan T and Anjewierden, Anjo and van den Berg, Ed and de Jong, Ton},
  journal={Physics Education},
  volume={55},
  number={6},
  pages={065026},
  year={2020},
  publisher={IOP Publishing}
}

@article{goorney2024framework,
  title={A framework for curriculum transformation in quantum information science and technology education},
  author={Goorney, Simon and Bley, Jonas and Heusler, Stefan and Sherson, Jacob},
  journal={European Journal of Physics},
  volume={45},
  number={6},
  pages={065702},
  year={2024},
  publisher={IOP Publishing}
}

@article{bungum2022quantum,
  title={What do quantum computing students need to know about quantum physics?},
  author={Bungum, Berit and Selst{\o}, S{\o}lve},
  journal={European Journal of Physics},
  volume={43},
  number={5},
  pages={055706},
  year={2022},
  publisher={IOP Publishing}
}

@article{michelini2023research,
  title={Research studies on learning quantum physics},
  author={Michelini, Marisa and Stefanel, Alberto},
  journal={The International Handbook of Physics Education Research: Learning Physics},
  pages={8--1},
  year={2023},
   url = {https://doi.org/10.1063/9780735425477_008},
 DOI = {10.1063/9780735425477_008},
  publisher={AIP Publishing LLC}
}

@article{schalkers2024explaining,
  title={Explaining {G}rover’s algorithm with a colony of ants: A pedagogical model for making quantum technology comprehensible},
  author={Schalkers, Merel A and Dankers, Kamiel and Wimmer, Michael and Vermaas, Pieter},
  journal={Physics Education},
  volume={59},
  number={3},
  pages={035003},
  year={2024},
  publisher={IOP Publishing}
}

@article{donhauser2024empirical,
  title={Empirical insights into the effects of research-based teaching strategies in quantum education},
  author={Donhauser, Anna and Bitzenbauer, Philipp and others},
  journal={Physical Review Physics Education Research},
  volume={20},
  number={2},
  pages={020601},
  year={2024},
  publisher={APS}
}

\newpage

\end{document}